\documentclass[a4paper,fleqn]{cas-dc}

\usepackage[authoryear]{natbib}

\def\tsc#1{\csdef{#1}{\textsc{\lowercase{#1}}\xspace}}
\tsc{WGM}
\tsc{QE}

\usepackage{tablefootnote}
\let\oldAA\AA
\renewcommand{\AA}{\text{\normalfont\oldAA}}

\begin{document}
\let\WriteBookmarks\relax
\def\floatpagepagefraction{1}
\def\textpagefraction{.001}

\shorttitle{Quantum Rate Theory and Electron Transfer Rate Dynamics}    

\shortauthors{Paulo R. Bueno}  

\title [mode = title]{Quantum Rate Theory and Electron-Transfer Dynamics: A Theoretical and Experimental Approach for Quantum Electrochemistry}  
                     
\author[1]{Paulo Roberto Bueno}[type=editor,auid=000,bioid=1, orcid=0000-0003-2827-0208]
                        
\cormark[1]

\ead{paulo-roberto.bueno@unesp.br}

\affiliation[1]{organization={Departmnet of Engineering, Physics and Mathematics, Institute of Chemistry, São Paulo State University}, 
            city={Araraquara},
            postcode={14800-060}, 
            state={São Paulo},
            country={Brazil}}


\begin{abstract}[summary]
Quantum rate theory is based on a first-principle quantum mechanical rate concept that comprises with the Planck-Einstein relationship $E = h\nu$, where $\nu = e^2/hC_q$ is a frequency associated with the quantum capacitance $C_q$ and $E = e^2/C_q$ is the energy associated with $\nu$. For a single state mode of transmittance, $e^2/C_q$ corresponds to the chemical potential differences $\Delta \mu$ between donor and acceptor state levels comprising an electrochemical reaction. A statistical mechanic treatment of $E$ is required to compute the contribution of the thermal dynamics at finite temperature. The Arrhenius equation for the temperature dependence of the reaction rate was obtained, as well as Marcus's Arrhenius-type electron-transfer rate constant as a particular setting of the quantum rate $\nu$. Consequently, this $\nu$ concept provides the quantum mechanical foundations for electrochemical reactions at room temperature. The present work also demonstrates that the electron-transfer rate of heterogeneous (diffusionless) reactions can be studied in detail within this theory by measuring $C_q$ using time-dependent electrochemical methods. Since the electron transfer follows a statistical mechanics version of the Planck-Einstein $E = h\nu$ relationship, the electrochemical reaction dynamics cannot be appropriately modeled using non-relativistic Schr\"odinger wave mechanics, which is the ongoing quantum approach to electrochemistry. Accordingly, a relativistic analysis that takes into account the spin dynamics of the electron is more appropriate. The latter assumption implies quantum electrodynamics within a particular quantum transport mode intrinsically coupled to the electron-transfer rate of electrochemical reactions that have not been considered thus far. Here it is demonstrated that the consideration of this inherent quantum transport is key to obtaining an in-depth understanding of the electron transfer phenomenon. Finally, the theory is validated through its description of electron transfer, quantum conductance, and capacitance in different electro-active molecular films.
\end{abstract}




\maketitle

\section{Introduction}\label{sec:introduction}

As a subset of the development of quantum mechanics in the field of chemistry~\citep{Levine-book}, quantum electrochemistry~\citep{Dogonadze-1970, Dogonadze-1972} is commonly known as the body of knowledge that arises from the use of non-relativistic Schr\"odinger wave quantum mechanics to the field of electrochemistry~\citep{Bard-book}, as a subset of the development of quantum mechanics to the field of chemistry~\citep{Levine-book}. In other words, the field of quantum chemistry refers to the application of quantum wave theory to predict the properties of atoms and molecules, and the use of this theory in electrochemistry is solely a subset of this field, wherein the electrodynamics of the electrochemical reactions are analyzed using the non-relativistic Schr\"odinger quantum equation. 

\begin{figure}[h]
\centering
\includegraphics[width=6cm]{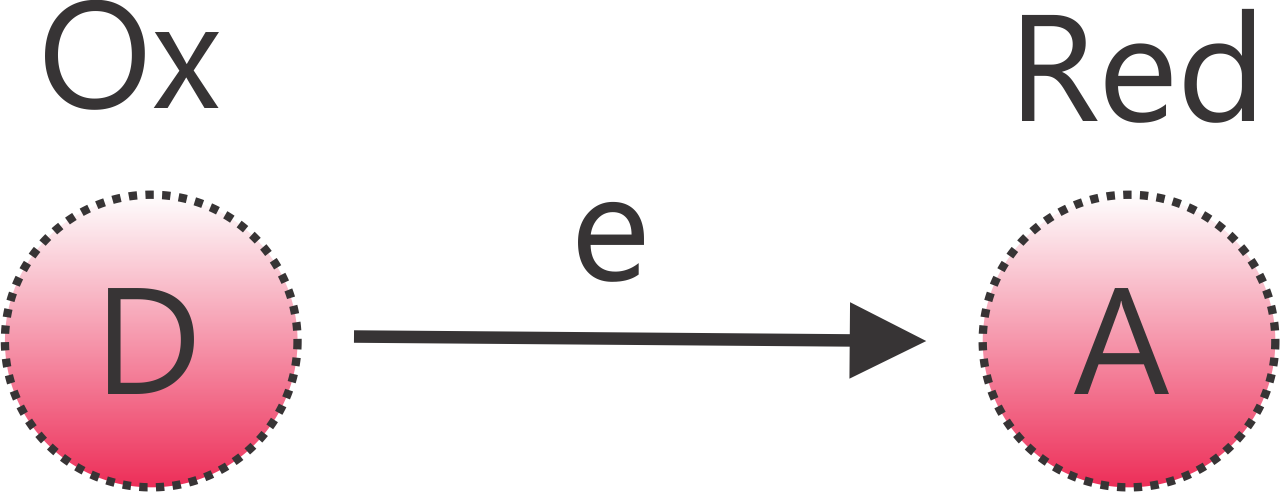}
\caption{Donor (oxidation) and Acceptor (reduction) states of an electrochemical reaction Ox + e $\rightharpoonup$ Red, otherwise known as a redox reaction. The present work will demonstrate that the dynamics governing electrochemical reactions follow relativistic quantum mechanics. Relativistic quantum mechanics can be described using a quantum resistive-capacitive circuit, which permits the investigation of the reaction dynamics with unprecedented experimental precision.}
\label{fig:D-A}
\end{figure}

Accordingly, quantum electrochemistry describes the dynamics of electron particles exchanged between the electronic states of the particular donor and acceptor ions or molecules during an electrochemical reaction. The molecules that receive electrons are referred to as acceptors (oxidizers), while those that donate electrons are donors (reducers). Understanding the dynamics of electron transfer from donor ($D$) to acceptor ($A$) states is the essence of the field of electrochemistry~\citep{Bard-book}, in which the associated chemical reaction is known as a redox (oxidation-reduction) reaction, as illustrated in Figure~\ref{fig:D-A}. 

There is no argument against the fact that the transfer of electrons between $D$ and $A$ states is essentially a quantum mechanical event owing to the intrinsic atomic scale on which it occurs. The rate at which this electron transfer (ET) dynamics between $D$ and $A$ states occurs is determined by the constant named ET rate constant $k$, constituting a key concept that governs the kinetics of electrochemical reactions and can be measured using well-known electrochemical methods~\citep{Bard-book} that employ electrodes to probe redox reactions.

The main purpose of the present work is to demonstrate that ET rate dynamics cannot be correctly modeled using Schr\"ordinger's non-relativistic quantum-wave methods, but it can be modeled if relativistic quantum electrodynamics methods are employed. These relativistic quantum electrodynamics can be suitably modeled using quantum resistive-capacitive (RC) circuit analysis within the quantum rate theory description of electrochemical reactions.

\section{Quantum Rate Theory}\label{sec:QrateT}

The quantum rate theory defines a fundamental quantum rate $\nu$ principle simply as the ratio between the reciprocal of the von Klitzing constant\footnote{This constant was named in honor of Klaus von Klitzing for his discovered of the quantum Hall effect, and is listed in the National Institute of Standards and Technology Reference on Constants, Units, and Uncertainty.} $R_k = h/e^2$ and the quantum (or chemical) capacitance, such as in~\cite{Bueno-2023-1, Bueno-2023-2}

\begin{equation}
 \label{eq:nu}
	\nu = \frac{e^2}{hC_q},
\end{equation}

\noindent where by noting that $E = e^2/C_q$ is an energy intrinsically associated with the electronic structure, it straightforwardly leads to the Planck-Einstein relationship, i.e.

\begin{equation}
 \label{eq:Planck-Einstein}
	E = h\nu = \hbar \textbf{c}_* \cdot \textbf{k},
\end{equation}

\noindent which follows a linear relationship dispersion between the energy $E$ and wave-vector\footnote{Note that \textbf{k} in bold refers to the wave-vector and its magnitude will be referred to as |\textbf{k}| to avoid notation issues with the meaning of $k$ that later will be referred to as the electron transfer rate constant.} $\textbf{k}$, where $\textbf{c}_*$ is the Fermi velocity and $\hbar$ is the Planck constant $h$ divided by $2\pi$. Hence, the quantum rate principle within Eq.~\ref{eq:nu} predicts relativistic dynamics that comply with Dirac instead of with the non-relativistic Schr\"ordinger equation, which can be verified by invoking the De Broglie relationship, which states that the momentum $\textbf{p} = \hbar \textbf{k}$ is directly related to the $\textbf{k}$ and hence Eq.~\ref{eq:Planck-Einstein} turns into $E = \textbf{p} \cdot \textbf{c}_*$, demonstrating its intrinsic relativistic character. 

\begin{figure}[h]
\centering
\includegraphics[width=7cm]{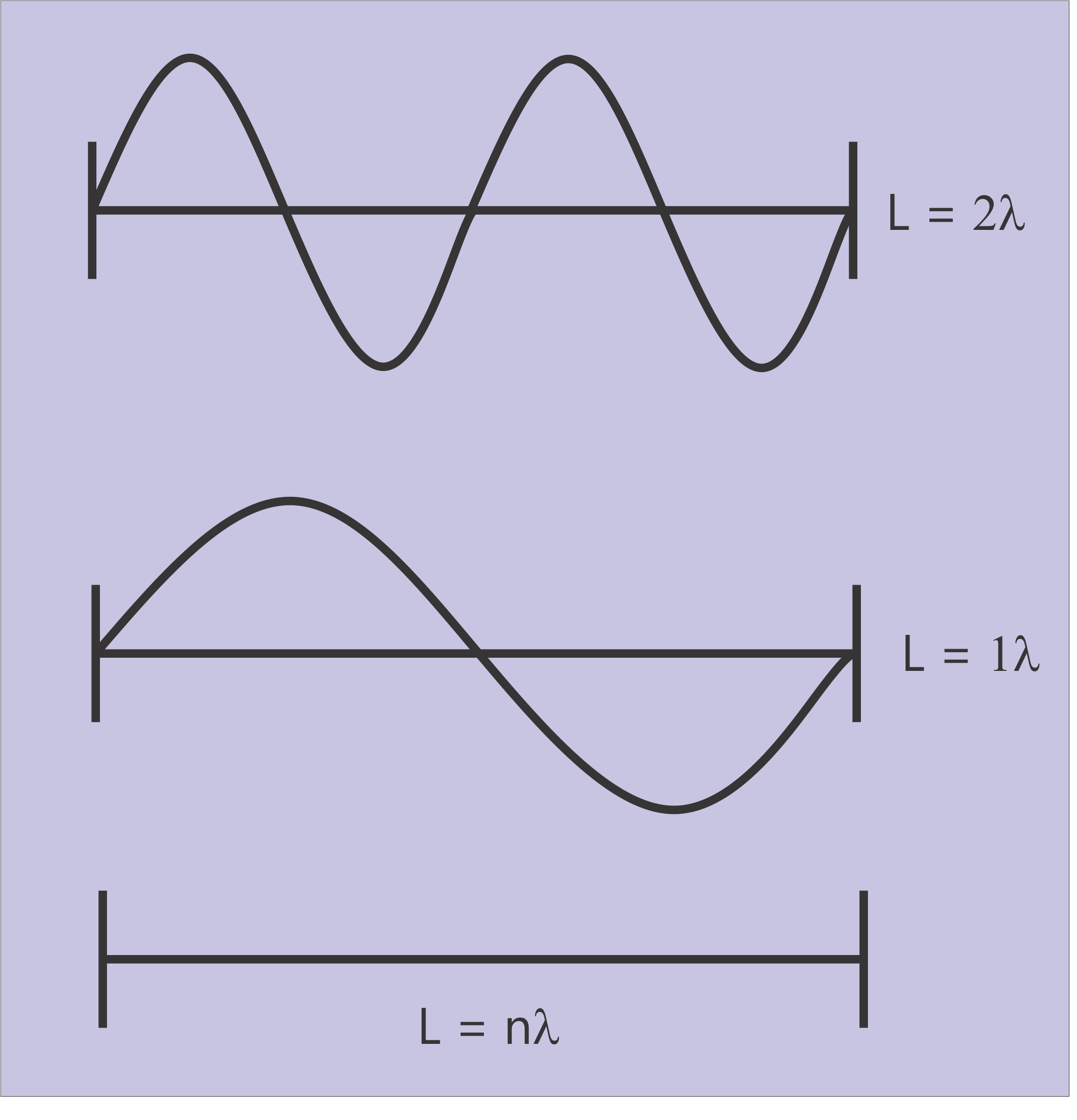}
\caption{Wave modes of electrons in quantum channels of length $L$ that comply with an integer number $n$ of wavelength $\lambda$, such that $L = n \lambda$ is obeyed.}
\label{fig:Qchannel}
\end{figure}

Therefore, whether the dynamics of electrons, during electrochemical reactions, follow the above-described Planck-Einstein relationship, as is the case to be demonstrated further, this implies that the electrodynamics of electrochemical reactions are governed by the Dirac (relativistic) instead of by the Schr\"odinger (non-relativistic) equation. If this is the assumption, it is important to account for electron spin dynamics to theorize the meaning of the ET rate constant. Consequently, the rate of the electron-transfer reactions is quantized, within a value of $G_0/C_q = g_sE/h$ for an adiabatic ideal quantum channel, where the electron-spin degeneracy is of key importance because it allows the modeling of the redox reaction using a quantum RC-circuit dynamics.

\section{Quantum Rate Theory and Electrochemistry}\label{sec:QrateT+EChem}

How does relativistic quantum mechanics can be invoked to model electrodynamics of electrochemical reactions? To answer this question let us first demonstrate how the electron spin dynamics can be incorporated to the quantum rate concept $\nu$ stated in Eq.~\ref{eq:nu} and further combine the results with the concept of quantum capacitance $C_q$.

\subsection{Conductance Quantum and Landauer Quantum Conductance}\label{sec:G0+Landauer}

The easiest way to incorporate the spin dynamics into Eq.~\ref{eq:nu} is by multiplying the spin degeneracy of the electron $g_s$ to the reciprocal of the von Klitzing constant $e^2/h$, which is expressed as follows:

\begin{equation}
 \label{eq:G_0}
	G_0 = g_s \frac{e^2}{h}.
\end{equation}

As a universally known constant, $G_0$ is referred to as the conductance quantum, possessing a value of $\sim$ 77.5 $\mu$S. The reciprocal of Eq.~\ref{eq:G_0}, i.e., $R_q = 1/G_0$, is the resistance quantum, which has a value of $\sim$ 12.9 k$\Omega$.

$G_0$ can be alternatively derived within the assumption of Eq.~\ref{eq:Planck-Einstein} and the additional consideration of a quantum channel governing the quantum dynamics of the electron transport between $D$ and $A$ states, according to~\cite{Bueno-2020}. The number $n$ of quantum wave modes within a quantum channel of length $L$, as illustrated Figure~\ref{fig:Qchannel}, is given by the ratio between $L$ and $\lambda$. For instance, considering that the electron is transported with the Fermi velocity $c_* = \lambda \nu$ in the channel of length $L = n \lambda$, Eq.~\ref{eq:Planck-Einstein} can be written as $E = h \left( nc_*/L \right)$, from which the derivative of $E$ concerning $n$, i.e. $\left( dE/dn \right)$, provides the density-of-states (DOS) per unit of length within the quantum channel as $\left( dn/dE \right) = 1 / c_*h$. 

\begin{figure}[h]
\centering
\includegraphics[width=6cm]{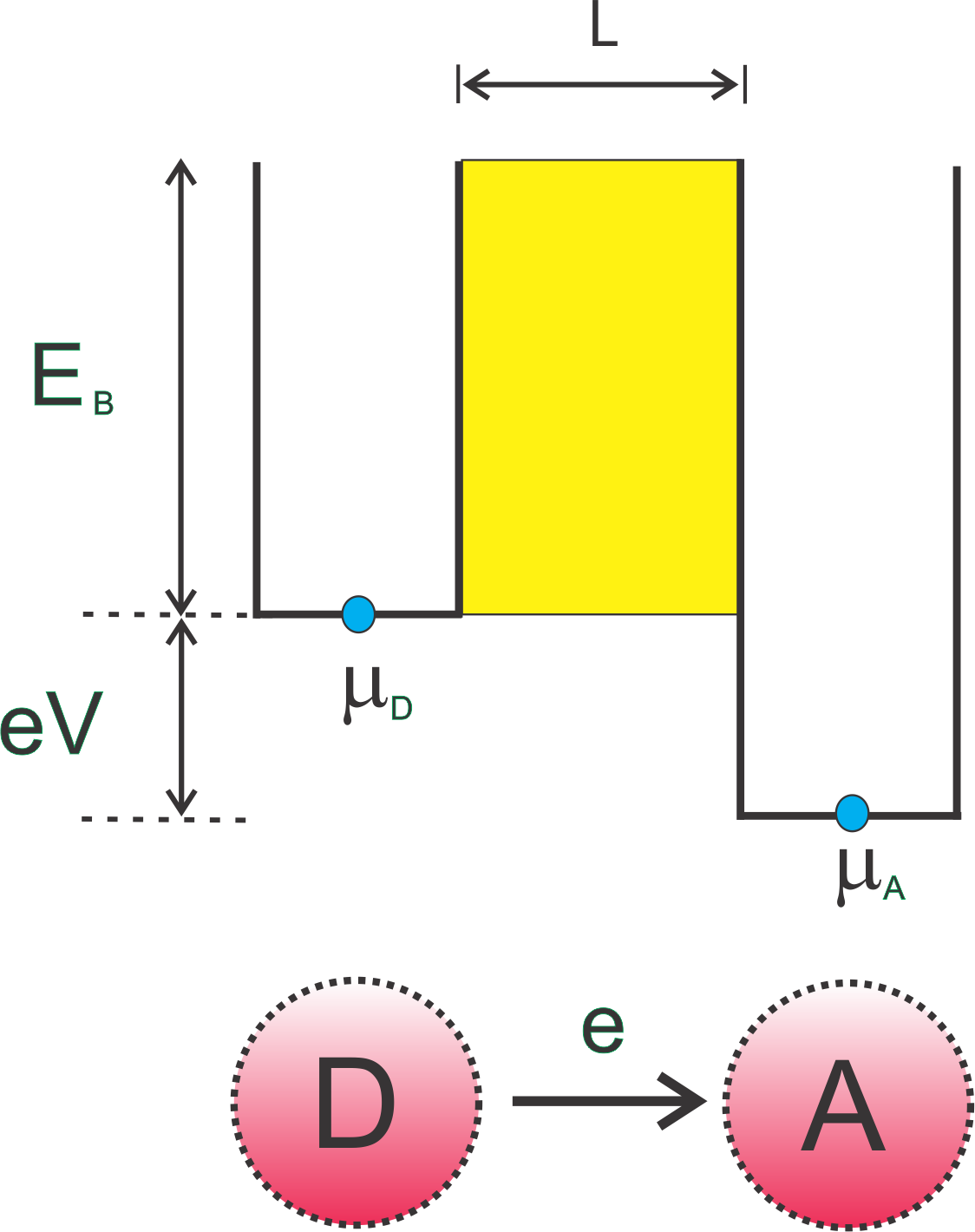}
\caption{$D$ and $A$ states separated by an effective distance of $L$ comprising a spin degenerate quantum channel within $n$ quantum modes for the transmittance of the electrons. $L$ corresponds to the length of the quantum channel, $E_{B}$ is the potential energy of the barrier that defines a tunneling electron coupling between $D$ and $A$ such that $\exp \left( {-\beta L} \right)$, known as the transmission coefficient, where $\beta$ is a constant that defines the properties of the barrier; and $\mu_{D}$ and $\mu_{A}$ are the chemical potentials states $D$ and $A$, defining a chemical potential difference that complies with $\Delta \mu = -eV = e^2/C_q$. In the particular case in which $E_{B}$ is zeroed, the quantum transport is adiabatic or ballistic with a perfect transmittance, and the conductance is defined theoretically as $G_0 = g_s e^2/h \sim$ 77.5 $\mu$S.}
\label{fig:DA-channel}
\end{figure}

Owing to the electric current per state requires a spin degeneracy of $g_s$ in the quantum channel of length $L$, which is defined as $L di/dn = -g_s e c_*$, from which an electric current per energy level and unit of length $L$ can be obtained as $di/dE = -g_s e c_*\left( dn/dE \right)$. Combining the latter result with $\left( dn/dE \right) = 1 / c_*h$, $di/dE = -g_s e / h$ is finally obtained. Now, to demonstrate that this results leads to $G_0$, it can be noted that $dE = -edV = \mu_D - \mu_A$, as illustrated in Figure~\ref{fig:DA-channel}, resulting that $G_0 = di/dV = g_s e^2/h$, \textit{quod erat demonstrandum}.

Since the magnitude of the wave-vector is $\textbf{|k|} = \left( 2\pi / \lambda \right)$ in Eq.~\ref{eq:Planck-Einstein}, it can be correlated with the length $L$ of the quantum channel by noting the number of quantum modes $n = L/\lambda$, as illustrated in Figure~\ref{fig:Qchannel}, permitting the expression of $\textbf{|k|}$ as a function of $n$ such that $\textbf{|k|} = n \left( 2\pi / L \right)$.

Eq.~\ref{eq:G_0} describes the ideal adiabatic conductance or resonance mode of electron transport in which electrons are transported through a perfect quantum channel with an ideal probabilistic transmittance that equates to the unity. 

Accordingly, generalities of Eq.~\ref{eq:G_0} to account, for instance, for non-adiabatic quantum electron transport modes, is required to a realistic consideration of the multiplicity of modes of quantum transport with a transmittance that differs from the adiabatic condition settled by $G_0$. In the latter situation, the electron transport no longer has ideal probabilities of transmittance. Hence, non-adiabatic situations can be modelled using scattering statistics, which can be accounted for by using a scattering matrix as an appropriate mathematical approach. Consideration of the transmission scattering matrix to model non-ideal electron transmittance is simply implemented by multiplying $G_0$ by a probability accounted by the scattering matrix such that

\begin{equation}
 \label{eq:Landauer}
	G \left(\mu\right) = g_s\frac{e^2}{h} \sum_{n=1}^{N}T_{n}\left( \mu \right),
\end{equation}

\noindent where $G \left(\mu\right)$ can be identified as a quantum of conductance accounted in a non-adiabatic regime of charge transport within a quantum channel with a defined transmittance lower than unity, known as the Landauer conductance~\citep{Landauer-1957}. Note that $G \left(\mu\right)$ is defined at a given chemical potential $\mu$ energy level, where $\sum_{n=1}^{N}T_{n}\left( \mu \right)$ is the transmission scattering matrix that states a transmittance probability through multiple $n$ channels, each possessing a transmittance of $T_{n}$ at a given $\mu$ energy-level state. $N$ in $\sum_{n=1}^{N}T_{n}\left( \mu \right)$ is the total number of quantum channels (or quantum modes within multiple channels) each possessing a transmittance of $T_{n}$ at a given energy level $\mu$.

The meaning of the conductance quantum $G_0$ and the quantum of conductance $G$ (as a multitude $G_0$ modes of transmission, each of them with a probability lower than the ideal unity situation) has now defined, and the meaning of $C_q$ it will be introduced in the next section.

\subsection{Quantum Capacitance and Electron Transfer}\label{sec:G0+Landauer}

The definition of capacitance is $1/C = dV/dq$, in which $dq = -e dn$. Provided that $d\mu = -edV$, as shown in the reaction in Figure~\ref{fig:DA-channel}, it can be straightforwardly demonstrated that $d\mu / dn = e^2/C_q$, which for $dn = 1$ provides the capacitance of a single electron transfer in a single-channel mode (as is the case depicted in Figure~\ref{fig:D-A}), which leads to $\Delta \mu = e^2/C_q$ for the quantum channels. Alternatively, note that $d\mu = dE/dn$, implying that the chemical potential difference between $D$ and $A$ states corresponds to the difference of energy $-eV$ between $D$ and $A$ states, as depicted in Figure~\ref{fig:DA-channel}. Owing to these two last above-mentioned assumptions, it can also be demonstrated that 

\begin{equation}
 \label{eq:Cq}
	C_q = e^2 \left( \frac{dn}{dE} \right),
\end{equation}

\noindent and hence the quantum capacitance $C_q$ (also referred to chemical capacitance~\citep{Bueno-book-2018}) is directly proportional to the density-of-state $\left( dn/dE \right)$~\citep{Bueno-book-2018, Bueno-2014-1, Luryi-1988}. For the specific situation depicted in Figure~\ref{fig:DA-channel}, the density-of-states (DOS) is equal to the number of quantum channels (or to the number of transmittance modes) per interval of energy that is available for the redox reaction to proceed.

In the next section, it will be demonstrated that the ratio between $G$ and $C_q$ provides a first-principle quantum mechanical meaning for the electron-transfer rate constant.

\subsection{Electron Transfer Rate and the Quantum of Conductance and Capacitance}\label{sec:G/C_q}

A key rate concept can now be defined~\citep{Bueno-2020, Bueno-book-2018}, which is given by the ratio of $G$ to $C_q$ such as

\begin{equation}
 \label{eq:k-zeroT}
	k = \frac{G}{C_q} = \frac{G_0}{C_q} \sum_{n=1}^{N}T_{n}\left( \mu \right) = g_s \nu \sum_{n=1}^{N}T_{n}\left( \mu \right),
\end{equation}

\noindent where $k$ is a frequency that is directly proportional to $\nu = E/h = e^2/hC_q$, as it was defined in Eq.~\ref{eq:nu}, based on the quantum rate concept.

\subsection{Equivalent Circuit Analysis of the Quantum Rate Dynamics}\label{sec:QR-circuit}

\begin{figure}[h]
\centering
\includegraphics[width=6cm]{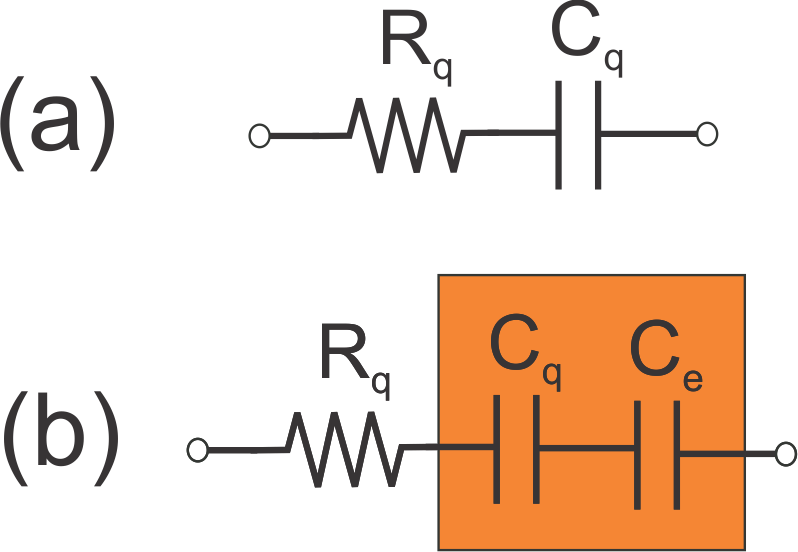}
\caption{(a) A redox reaction, such as that depicted in Figure~\ref{fig:D-A}, can be represented in terms of a quantum RC circuit comprised of a quantum resistor $R_q = 1/G$ and of a quantum capacitance $C_q$. (b) A more complex scenario involving a redox reaction coupled to an electrode, as depicted in Figure~\ref{fig:DA-electrode-electrolyte}. Here, the series resistance $R_s = R_c + R_e$, which is the sum of the contact $R_c$ and solution/electrolyte $R_e$ resistances, was omitted for the convenience of putting emphasis in the meaning of $C_\mu$, the equivalent capacitance $\left( 1/C_\mu = 1/C_e + 1/C_q \right)$ of the ET dynamics, as will be better introduced in section~\ref{sec:ElectroCap}. Nonetheless, $R_s$ is an important practical component of the dynamics and will be properly discussed in sections~\ref{sec:Meaning-DOS} and \ref{sec:QRct}.}
\label{fig:QRC-circuit}
\end{figure}

From Eq.~\ref{eq:k-zeroT}, it can be observed that $G = 1/R_q = G_0 \sum_{n=1}^{N}T_{n}\left( \mu \right)$ in a way that $k = 1/\tau$, where $\tau = R_q C_q$ is the characteristic time constant associated with the redox reaction dynamics, as it was previously depicted in Figure~\ref{fig:D-A}. This corresponds to a series quantum RC circuit, as represented in Figure~\ref{fig:QRC-circuit}. 

Note that the quantum RC circuit depicted in Figure~\ref{fig:QRC-circuit} applies to both adiabatic and non-adiabatic ET processes. For an adiabatic situation, in which the transmittance is ideal, $\sum_{n=1}^{N}T_{n}\left( \mu \right)$ is the total number of quantum modes $N$. For a single electron transport mode within of a single channel, $N$ reduces to the unit and $k = 1/\tau = g_s e^2/hC_q$, which is the situation stated in Eq.~\ref{eq:nu}, with a spin degeneracy of $g_s$ for $E$, i.e., for $E = g_s e^2/C_q$.

The RC dynamics depicted in Figure~\ref{fig:QRC-circuit} can be investigated using impedance-derived capacitance spectroscopy, in which the complex capacitance can be defined as~\cite{Bueno-book-2018}

\begin{equation}
 \label{eq:Complex-Cq}
	C^{*}(\omega) = \frac{C_q^0}{1 + j\omega \tau} \sim C_q^0 \left(1 - j\omega \tau \right).
\end{equation}

The term $C_q^0$ in Eq.~\ref{eq:Complex-Cq} corresponds to the equilibrium capacitance which is obtained for $\omega \rightarrow 0$, where the imaginary component of the complex capacitance is negligible.

In the next section, the thermodynamics of the quantum rate are discussed.

\subsection{The Thermodynamics of the Quantum Rate}\label{sec:Therm-QR}

The energy $E = e^2/C_q = h \nu$ term of Eq.~\ref{eq:k-zeroT} is not consistent with the thermodynamics of the process because it is formulated at the zero-temperature limit. In other words, the temperature dependence required for a room temperature analysis of the ET dynamics of electrochemical reactions is not evaluated in Eq.~\ref{eq:k-zeroT}. 

The thermodynamics that conducts to the thermal broadening of $E = e^2/C_q$ can be quantified using statistical mechanics. Particularly, using the grand canonical ensemble presumption, it can be demonstrated that $E = e^2/C_q = k_BT/N \left[f(1-f) \right]^{-1}$, where $f = \left(1 + \exp  \left(E / k_{B}T \right) \right)^{-1}$ is the Fermi-Dirac distribution function, $k_B$ is the Boltzmann constant and $T$ is the absolute temperature. Within statistical mechanics considerations, Eq.~\ref{eq:k-zeroT} can be rewritten as

\begin{equation}
 \label{eq:k-finiteT}
	k = \frac{G}{C_q} = G_0 \left( \frac{k_BT}{e^2N} \right) \left[f(1-f) \right]^{-1} \sum_{n=1}^{N}T_{n}\left( \mu \right).
\end{equation}

It can be observed that Eq.~\ref{eq:k-finiteT}, besides comprising non-adiabatic processes, also incorporates the required statistical mechanics considerations of the quantum mechanics stated in Eq.~\ref{eq:Planck-Einstein}. In other words, the non-adiabatic considerations of the quantum transport of electrons of Eq.~\ref{eq:Landauer} from Eq.~\ref{eq:G_0} are taken into account by introducing $\sum_{n=1}^{N}T_{n}\left( \mu \right)$ scattering statistics. This introduction of $\sum_{n=1}^{N}T_{n}\left( \mu \right)$ permits the modelling of the electron coupling between $D$ and $A$ states in a multitude of ways. 

The use of the transmission scattering matrix $\sum_{n=1}^{N}T_{n}\left( \mu \right)$ approach instead of the most traditional method that employs the donor-acceptor electronic coupling matrix $|H_{DA}|$~\citep{Zhu-2021} is advantageous because it permits a direct comparison with the quantum transport observed in molecular and nanoscale electronics~\citep{Santos-2020}, where quantum rate theory also applies to the study of ET dynamics~\citep{Santos-2020}. 

For instance, in the specific situation that quantum transport of electrons is conducted through a potential barrier governed by tunneling, the term $\sum_{n=1}^{N}T_{n}\left( \mu \right)$, in equation Eq.~\ref{eq:k-finiteT}, equates to the transmission coefficient $\kappa = \exp \left( {-\beta L} \right)$, where $\beta$ is a parameter associated with the properties of the barrier with a length $L$ for electrons to tunnel, as depicted in Figure~\ref{fig:DA-channel}. In the latter case, the effective distance between $D$ and $A$ states is the length $L$ of the potential barrier with an energy $E_B$.

According to the above-introduced assumption, where the mechanism of electron transfer between $D$ and $A$ follows a tunneling dynamics, the Landauer quantum conductance takes the form $G = \kappa G_0N = \left( \kappa e^2N/h \right)$ in Eq.~\ref{eq:k-finiteT} and owing to

\begin{equation}
 \label{eq:Cq-thermal}
	C_{q} = \left( \frac{e^2N}{k_{B}T} \right) \left[ f(1-f) \right],
\end{equation}

\noindent the frequency $k$, in Eq.~\ref{eq:k-finiteT}, defined as $G/C_{q}$, can now be written as~\citep{Bueno-2020}

\begin{equation}
 \label{eq:Marcus-quantum}
	k = \kappa \frac{k_BT}{h} \left[ f\left(1-f \right) \right]^{-1} = \kappa \frac{k_BT}{h} f^{-2} \exp \left( E /k_BT \right),
\end{equation}

\noindent which turns into the Arrhenius format

\begin{equation}
 \label{eq:Arrhenius}
	k = \kappa \frac{k_BT}{h}\exp \left(- \frac {E} {k_{B}T} \right),
\end{equation}

\noindent whenever the Boltzmann approximation is applied to Eq.~\ref{eq:Marcus-quantum}, which is the case where $1 << \exp \left( E /k_{B}T \right)$ in $f = \left(1 + \exp  \left(E / k_{B}T \right) \right)^{-1}$.

As Marcus ET theory~\citep{Marcus-1964, Marcus-1985, Sumi-1986, Marcus-1993} is derived from transition state theory (TST), which is a particular situation in the Arrhenius equation stated in Eq.~\ref{eq:Arrhenius}, it is clear that the Marcus ET theory is a particular setting of the quantum rate theory, as demonstrated here from the first-principles relativistic quantum mechanical concepts within Eq.~\ref{eq:nu}. 

In TST, the activation energy $E$ in Eq.~\ref{eq:Arrhenius} is replaced by $E^{\ddagger}$, which considers a quasi-equilibrium condition in which $E^{\ddagger}$ is the energy associated with the activated complex, an intermediate and `excited' energy state achieved by the reactants, before the reaction proceeds from the reactants (Ox + $e$) to the products (Red). 

Marcus's contribution~\citep{Marcus-1964} was to consider that electron transfer occurs predominately through the energy $E^{\ddagger}$ activated step (see Figure~\ref{fig:Qrate-SC-ET}), but in a way that the effect of the solvent environment is fundamental. In other words, the energy of the activated complex, i.e. $E^{\ddagger}$, is written as a function of an additional energetic term known as the reorganization energy of the solvent $\lambda_0$. Accordingly, Marcus's original ET theory predicts that $E^{\ddagger} = \left( -eV + \lambda_0 \right)^2 / 4\lambda_0$, where $-eV$ is accounted as the difference between $D$ and $A$ chemical potentials in the quantum rate nomenclature, as indicated in Figure~\ref{fig:Qrate-SC-ET}. 

Therefore, by taking this key Marcus ET rate assumption into Eq.~\ref{eq:Arrhenius} and generalizing the electron coupling using transmission scattering matrix $\sum_{n=1}^{N}T_{n}\left( \mu \right)$ in place of $|H_{DA}|$, ET reactions, in homogeneous settings (as is the original proposal of Marcus ET theory), is written in terms of its derivation from the quantum rate theory as

\begin{equation}
 \label{eq:Arrhenius-Marcus}
	k = \frac{k_BT}{h} \frac{\sum_{n=1}^{N}T_{n}\left( \mu \right)}{\sqrt{4\pi\lambda_0k_{B}T}} \exp \left[- \frac {\left( -eV + \lambda_0 \right)^2 } {4\lambda_0k_{B}T} \right],
\end{equation}

\noindent which is a normalized Gaussian function and is also known as the Marcus's DOS~\citep{Bard-book}. Note that Eq.~\ref{eq:Arrhenius-Marcus} is a slightly different way of depicting the traditional Marcus's DOS, in which $\sum_{n=1}^{N}T_{n}\left( \mu \right)$ is suitably considered in place of the donor-acceptor electronic coupling matrix $|H_{DA}|$. 

The use of $\sum_{n=1}^{N}T_{n}\left( \mu \right)$ instead of $|H_{DA}|$ has advantages because $\sum_{n=1}^{N}T_{n}\left( \mu \right)$ is intrinsically related to quantum channels, which is a fundamental constitutive concept of the quantum rate theory that allows a direct correlation between ET and quantum transport in a relativistic quantum mechanical description of the event, demonstrating that electrochemistry and molecular electronics have the same quantum-mechanics foundations. 

In other words, the presence of a quantum conductance intrinsically involved with the homogeneous ET dynamics has so far been ignored, although it can be incorporated into the traditional ET semi-classical theory. More details will be provided in the next section.

\subsection{Quantum Transport within Semi-Classical Electron Transfer Theory}\label{sec:QTrans-SC-ET}

\begin{figure}[h]
\centering
\includegraphics[width=8cm]{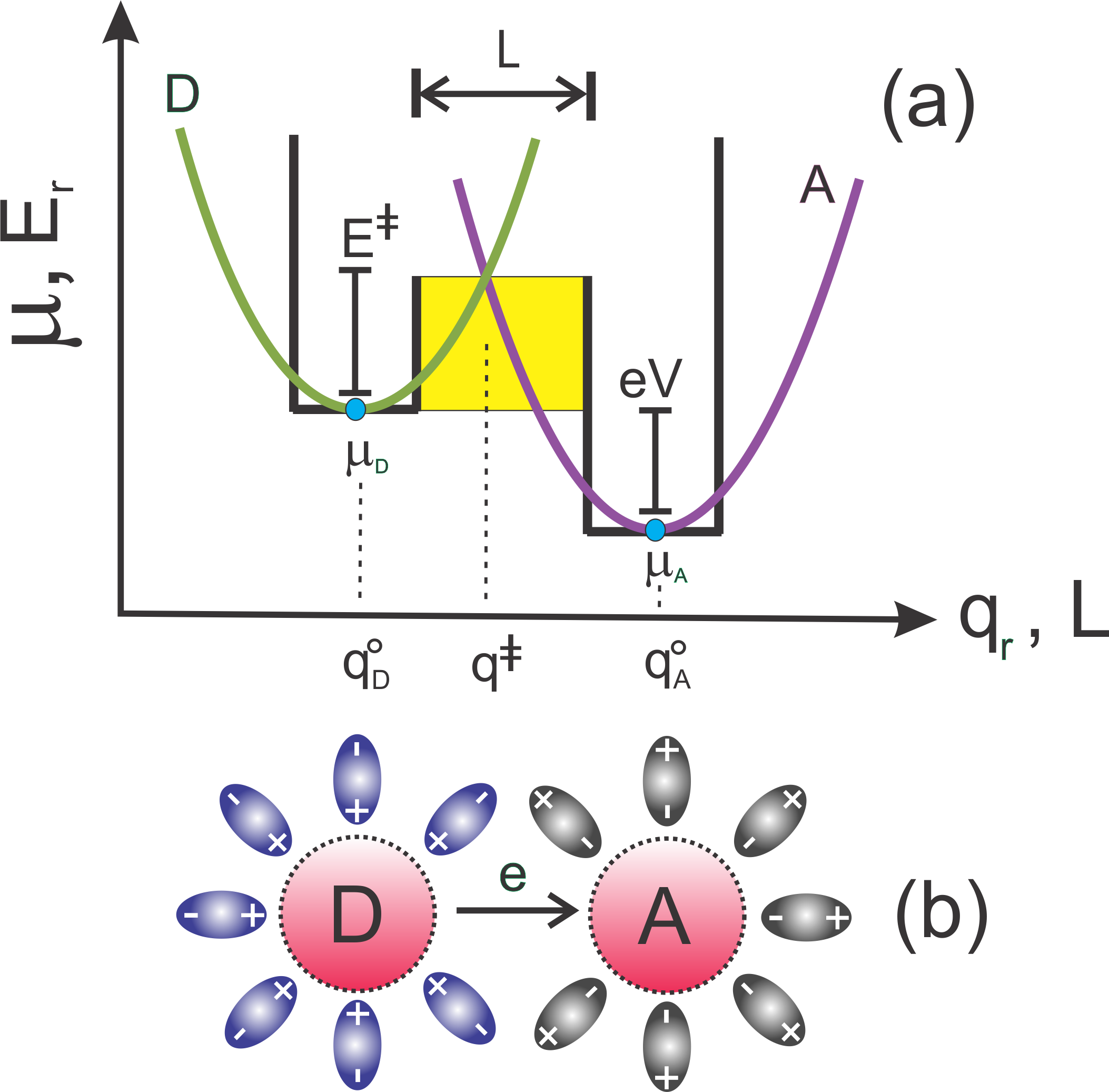}
\caption{(a) Superposition of rectangular and parabolic potential wells. As discussed in the text, the quantum rate theory can describe the electron rate dynamics independently of the internal structure of $D$ and $A$. The latter assumption arise because the electronic structure of the channel is contained in the definition of $C_q = e^2 \left( dn/dE \right)$, which is proportional to the electronic DOS of the channel. (b) Representation of $D$ and $A$ structures under the influence of the solvent environment. The solvent environment is an additional dynamics that is computed in Eq.~\ref{eq:Arrhenius} ($E$ is replaced by $E^\ddagger$), which was deduced from the approximation of Eq.~\ref{eq:Marcus-quantum} to the Boltzmann statistics. Along the ordinate axis, it is depicted as the free-energy $E_r$ (chemical potential is the equivalent nomenclature adopted by the quantum rate theory) of the reaction is plotted, while the abscissa axis is the reaction coordinate $q_r$ (corresponding to the quantum channel length $L$ in nomenclature adopted by the quantum rate theory).}
\label{fig:Qrate-SC-ET}
\end{figure}

It will be now demonstrated that the internal structure of $D$ and $A$ in Figure~\ref{fig:Qrate-SC-ET}\textit{b} can be depicted either as parabolic or rectangular potential shapes without influencing the derivation of Eq.~\ref{eq:Arrhenius} or Eq.~\ref{eq:Arrhenius-Marcus}. The latter is a consequence of the meaning of $C_q$, as introduced in Eq.~\ref{eq:Cq}, which contains information concerning the internal $D-A$ adjoined solvating structure formed during the ET event in a homogeneous or heterogeneous setting. The term $e^2/C_q = \Delta \mu$ is $-eV$, corresponding to the variation in chemical potential per unit of electron particles as the Gibbs-free energy of a single ET process, as indicated in Figure~\ref{fig:D-A}. Therefore, $e^2/C_q$ accounts for the superposition of the frontier orbitals of the $D-A$ structure in a Density-Functional Theory (DFT) definition of $e^2/C_q$~\citep{Bueno-2017-2, Miranda-2019, Bueno-book-2018}.

Important to be accounted for to determine the reaction dynamics is the number of quantum channels that establish the number of quantum modes available for the electron to be transferred from $D$ to $A$. This information can be expressed as the number density of quantum channel models (or quantum states) which is given by the density-of-states $\left( dn/dE \right)$.

Using the above definition of $\left( dn/dE \right)$ and noting that it correlates with Eq.~\ref{eq:Cq}, the correctness of this analysis of the ET dynamics can be demonstrated. In other words, Eq.~\ref{eq:k-zeroT} can be recovered by noting that $e^2(dn/dE)$ is $C_q$ in the analysis of Figure~\ref{fig:Qrate-SC-ET}. This corresponds to considering $g_se^2/C_q = \left( dE/dn \right) = h \left( c_*/L \right)$, where the spin degeneracy $g_s$ was introduced in the analysis to maintain the correlation with the relativistic dynamics. The relativistic frequency $\nu$ is $c_*/L$ frequency at which electrons cross the barrier of length $L$ with Fermi velocity of $c_*$. The characteristic time of the transit is $\tau = R_qC_q$, such as that $c_* = L/\tau$, straightforwardly demonstrating that $k = G/C_q$ is obtained as a result of the meaning of $C_q$. Note the straight correlation with the Planck-Einstein relativistic equation (Eq.~\ref{eq:nu}), which were ultimately in agreement with Eq.~\ref{eq:k-zeroT}. The conclusion of this section is that a quantum conductance $G$ is an intrinsic part of ET dynamics even for homogeneous setting.

In the next section, the meaning of the quantum capacitance is generalized and the meaning of electrochemical capacitance is introduced.

\section{General Definition of Quantum Capacitance}

In the previous section, it was demonstrated that the energy $E = -eV$ associated with $D$ and $A$ electrodynamics that comprises redox reactions is $E = e^2/C_q = \left( dE/dn \right)$, where $\left( dE/dn \right)$ is the amount of energy per amount of quantum channel modes within the $D-A$ complex. Hence, the quantum modes of electron transmittance are formed as a result of the formation of the $D-A$ complex during the occurrence of the ET event. The latter analysis directly leads to $\mu = dE/dn = -eV$, where $\mu = dE/dn$ is identified as the electronic chemical potential~\citep{Miranda-2016, Miranda-2019}, which is the free energy per number of elementary charge $e$ required for ET to proceed through a quantum channel of length $L$. This free energy per electron directly correlates with the conductance quantum associated with the redox reaction. The above analysis is in agreement with semi-classical homogeneous ET theory, as indicated in Figure~\ref{fig:DA-channel}, where $E^\ddagger$ is an additional contribution that accounts for the reorganization energy $\lambda_0$ of the solvent (solvating structure rearrangement) during ET, as it was originally proposed by Marcus.

\begin{figure}[h]
\centering
\includegraphics[width=8cm]{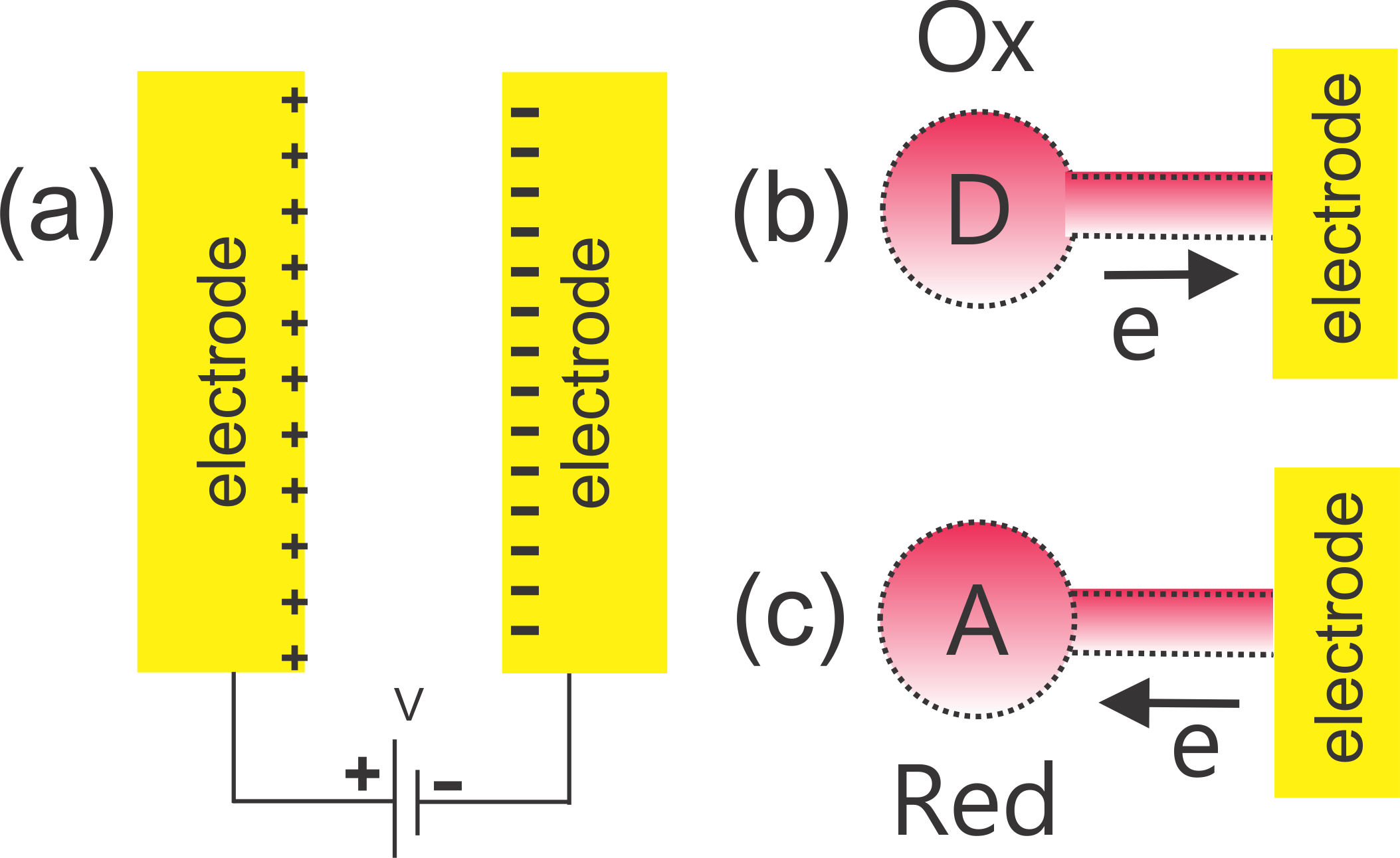}
\caption{(a) A quantum capacitive $C_q$ contribution is not present in a Coulombic microscopic separation of charge simply because the DOS of each plate of a classical capacitor is too high according to the definition provided in Eq.~\ref{eq:Cq-general}. In the case of a (b) $D$ and (c) $A$ molecular structures attached to an electrode plate serving as a probe, the DOS of $D$ and $A$ are accessible, which is also in agreement with the definition of $C_q$ provided by Eq.~\ref{eq:Cq-general}.}
\label{fig:DA-electrode}
\end{figure}

In this section, the general definition of the quantum capacitance $C_q$ will be discussed. The analysis to be conducted is important for understanding the origin of non-faradaic and faradaic electric current modes of charging electrochemical interfaces. For instance, it will be shown that faradaic currents are associated with changing quantum capacitive states, whereas non-faradaic currents are associated with a polarization (related to the dielectric characteristic of the environment) mode of charging electrochemical interfaces. Non-faradaic modes of charging electrochemical interfaces correspond to phenomena such as double-layer capacitance, in particular. 

Consequently, it will be concluded that faradaic currents are the origin of the pseudo-capacitive phenomena observed at particular electrochemical interfaces. An in-depth molecular scale analysis of this phenomenon~\citep {Bueno-2019} can enhance the development of supercapacitors and batteries, for instance. Electrolytic supercapacitors are based on non-faradaic electric charging modes, whereas hybrid supercapacitors take advantage of these two different capacitive charging modes to operate~\citep{Bueno-2019}.

The general definition of quantum capacitance depends on identification of the quantum contribution of two different reservoirs that are contacted. Taking the individual electronic structure of $D$ and $A$ separately without establishing, in principle, ET electrodynamics between both of these states, the quantum capacitance is defined as
 
\begin{equation}
 \label{eq:Cq-general}
	\frac{1}{C_q} = \frac{1}{e^2} \left[ \frac{1}{\left( dn/dE \right)_D} + \frac{1}{\left( dn/dE \right)_A} \right],
\end{equation}

\noindent where $\left( dn/dE \right)_D$ and $\left( dn/dE \right)_A$ are the DOS of $D$ and $A$ reservoirs, respectively.

If the general $C_q$ concept stated in Eq.~\ref{eq:Cq-general} is applied to the separation of an amount of charge $q$ over a potential difference $V$ applied to two metallic plate electrodes, as is the case represented in Figure~\ref{fig:DA-electrode}\textit{a}, it leads to $e^2/C_q$ = [1/DOS$_l$ + 1/DOS$_r$], where DOS$_l$ and DOS$_r$ are the DOS of the left and right metallic plates, respectively. 

As both DOS$_l$ and DOS$_r$ are very high, the energy contribution associated with the occupancy of states computed as $e^2/C_q$ tends to be null. In other words, there is no additional energy contribution besides that associated with the electrostatic coulombic spatial separation of charge between the plates of the capacitor.

\begin{figure}[h]
\centering
\includegraphics[width=7cm]{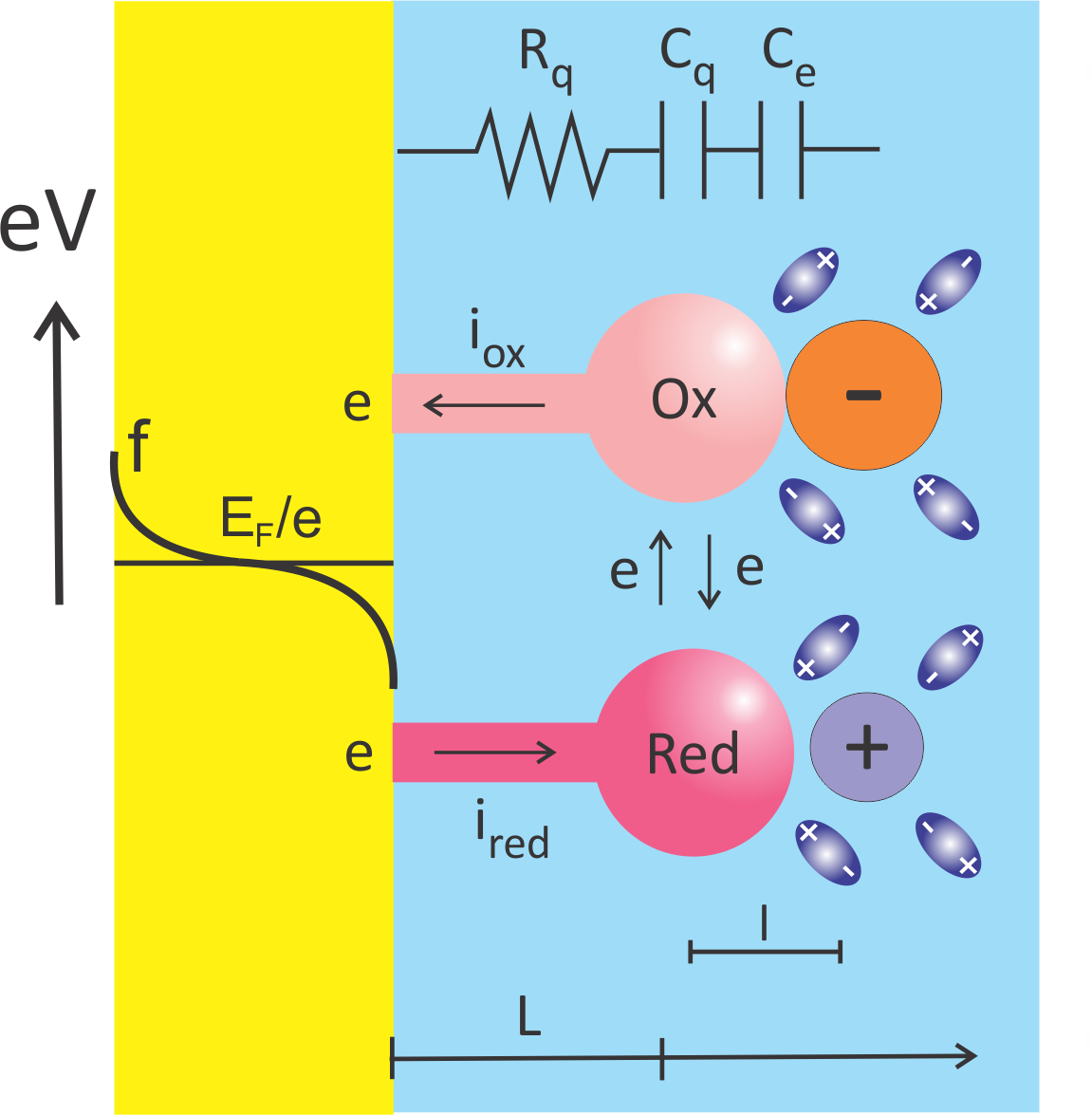}
\caption{Redox reaction comprised of a hypothetical redox pair covalently attached to an electrode (yellow) in an electrolyte environment (blue). $L$ represents the distance of separation of the redox pair is separated from the electrode comprising a potential barrier. In a tunneling mechanism of electron transport, electron coupling is modeled by a transmission coefficient such as that $\kappa \sim \exp \left( -\beta L \right)$. Following the electrodynamics predicted by Eq.~\ref{eq:Marcus-quantum}, the equivalent circuit that models this interface is that of Figure~\ref{fig:QRC-circuit}\textit{b}, in which $C_e$ is an additional series contribution associated with a spatial charge separation with a length $l$. The electric neutrality within $l$ is achieved by taking the contribution of the counter ions of the electrolyte, as shown in this figure. Note that the redox reaction, within reduction $i_{red}$ and oxidation $i_{ox}$ electrochemical currents, is accessible from the electrode at the formal potential of the redox pair that occurs at $E_F/e$. The Fermi-Dirac occupation statistics of the electron energies at the interface of this electrode are denoted by $f$.}
\label{fig:DA-electrode-electrolyte}
\end{figure}

Let us now analyze what occurs if $C_q$, as stated in Eq.~\ref{eq:Cq-general}, is applied to the situation of Figure~\ref{fig:DA-electrode}\textit{b} and Figure~\ref{fig:DA-electrode}\textit{c}, where, respectively, there is a molecular $D$ or $A$ state covalently coupled to an electrode. The analysis will lead to $e^2/C_{q,D} =$ $[1/\left( dn/dE \right)_D$ + 1/DOS$_e$] for $D$ and $e^2/C_{q,A} = $ $[1/\left( dn/dE \right)_A$ + 1/DOS$_e$] for $A$, where DOS$_e$ is the DOS of the electrode. Following similar arguments stated in the analysis of the previous paragraph, the term 1/DOS$_e$ tend to be null (owing to high DOS$_e$) for computing the capacitive contributions of $D$ and $A$, thus $C_{q,D} = e^2 \left( dn/dE \right)_D$ is obtained for the case of $D$ (Figure~\ref{fig:DA-electrode}\textit{a}) and $C_{q,A} = e^2 \left( dn/dE \right)_A$ for the case of $A$ (Figure~\ref{fig:DA-electrode}\textit{b}).

In the presence of an electrolyte and a redox-active molecular film attached to an electrode, a more realistic picture is that depicted in Figure~\ref{fig:DA-electrode-electrolyte}, which, in order to be understood in depth, requires an analysis of the electrochemical capacitance $C_\mu$, which will be discussed in detail in the next section.

\section{Electrochemical Capacitance and Modes of Charging Electrochemical Interfaces}\label{sec:QCap+Charging}

In the previous section, the general definition of quantum capacitance was introduced, and how it hypothetically applies to the case of $D$ and $A$ molecules individually coupled to an electrode was discussed (it was shown to act as an electron reservoir with a high DOS). In the present section, the concept of the quantum capacitance will be combined with the concept of classical coulombic capacitance. 

The combination will give rises to the concept of electrochemical capacitance, which is more appropriate for dealing with electrochemical interfaces comprising quantum molecular moieties attached to an electrode, which in its turn, is embedded into an electrolytic environment. The contact of a chemically modified electrode with an electrolyte allows an investigation of the electronic structure of the individual molecular moieties in the presence of an electron-ion pair, which represents a specific type of a nanoscale electric-field screening phenomenon~\citep{Pinzon-2022}.

\subsection{Electrochemical Capacitance}\label{sec:ElectroCap}

The concept of electrochemical capacitance $C_\mu$ is an equivalent capacitance that is computed as a result of a series combination of coulombic $C_e$ spatial separation of charge (also referred as polarization) and quantum $C_q$ capacitances, as it was defined in Eq.~\ref{eq:Cq-general}, such as that

\begin{equation}
 \label{eq:C-electroch}
	\frac{1}{C_\mu} = \frac{1}{C_e} + \frac{1}{C_q}.
\end{equation}

In the classical analysis of two capacitors arranged in series, in which only the coulombic charge separation operates, there is an equivalent amount of charges distributed between the two capacitors and a different electric potential in each of them. However, given the atomic and molecular scales of electrochemical capacitors, one of the capacitances has a different physical origin (unrelated to the Coulomb or Gauss law); however, they can have the same charge and potential difference. The consequences of this will be better discussed further as an electrochemical degeneracy $g_r$.

Worthy to note is the additional contribution of $C_e$ to $C_q$ in describing the ET dynamics. Hence, $C_e$ must be taken into account for describing the properties of the electrolyte structure within the contributions of the solvent and counter-ions to the quantum electrodynamics described by Eq.~\ref{eq:nu}. Adding the electrolyte contribution to the quantum rate dynamics, as predicted by Eq.~\ref{eq:k-zeroT}, it leads to~\citep{Bueno-book-2018}

\begin{equation}
 \label{eq:k-Celect-zeroT}
	k = \frac{G}{C_\mu} = G_0 \sum_{n=1}^{N}T_{n}\left( \mu \right) \left( \frac{1}{C_e} + \frac{1}{C_q} \right).
\end{equation}

Eq.~\ref{eq:k-Celect-zeroT} is a key equation because it allows us to predict non-faradaic $k_{nf} = G/C_e$ and faradaic modes $k = G/C_q$ of charging the electrochemical interfaces, as will be discussed in the next section.

\subsection{Faradaic and Non-faradaic Modes of Charging Electrochemical Interfaces}\label{sec:ChargingElecInterfaces}

In the present work, molecular redox films serve as models for studying the relativistic quantum dynamics of ET reactions. The traditional semi-classical interpretation of the ET dynamics of redox-molecular films is well-described in a review by Eckermann et al.~\citep{Eckermann} and is not elaborated on here.

In the absence of accessible redox molecular quantum states in an electrochemical interface comprising an electrode modified with a molecular layer or in a redox-active monolayer~\citep{Eckermann} poised in a bias potential (where there is no redox Faradaic activity and electrons cannot be exchanged with the electrode), Faradaic dynamics will be absent. This latter assumption means that there will not be a $C_q$ capacitive contribution to the response of a time-dependent energy perturbation of the interface. 

This is equivalent to settling $k_{nf} = G_{nf}/C_e$ in Eq.~\ref{eq:k-Celect-zeroT}, which is based on the consideration of $1/C_q$ as null. This physical situation corresponds to the well-known polarization charging of a capacitive interface where $C_e$ will be the dominant capacitance of the interface. A typical example is the measurement of the double-layer $C_{dl}$ capacitance of a metal interface in direct contact with an electrolyte. The time-dependent relaxation dynamics of interfaces in which $C_q$ is absent and the $e^2/C_q$ energy is inaccessible, respond differently to interfaces in which $C_q$ is present. The capacitive response of polarizable interfaces is well-known to follow time-dependent \textit{dielectric} relaxation dynamics~\citep{Goes-2012}. In summary, in the absence of $C_q$ in Eq.~\ref{eq:k-Celect-zeroT}, only $C_e$ responds to a time-dependent energy perturbation and there is a $k_{nf}$ that governs the charge dynamics of the interface, but having a different meaning of that of $k$ that follows a relativistic quantum dynamics aligned with Eq.~\ref{eq:nu}. 

\begin{figure}[h]
\centering
\includegraphics[width=6cm]{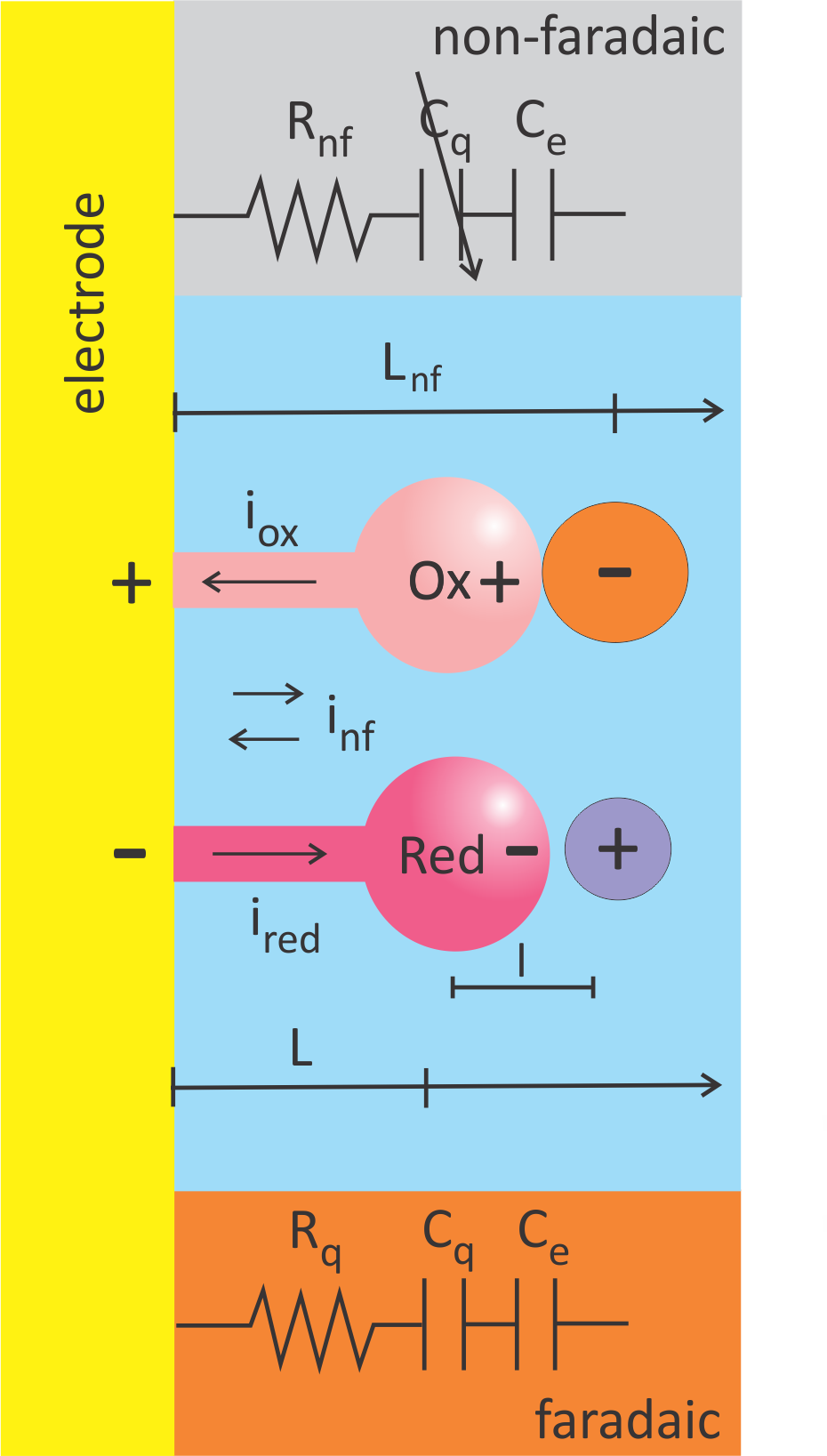}
\caption{This figure illustrates the charge dynamics within an electrochemical interface comprising two rate modes of charging the capacitive states of the interface, i.e. a non-faradaic polarization mode and a faradaic that is associated with charging quantum states. The non-faradaic mode of charging the interface is governed by a spatial charge separation within a length $L_{nf}$ that is proportional to the Debye length. This mode of charging the interface has an electrochemical current that is $i_{nf} = C_e s$. Additionally to this non-faradaic polarizing mode of charging the interface, there is a faradaic mode that is governed by an electrochemical current with a magnitude of $i = C_q s$. The length scale associated with $i$ is $L$, which accounts for the lengths of the quantum channels, as shown in Figure~\ref{fig:DA-channel}. It is important to note that in the case of the faradaic rate dynamics, there is a $e^2/C_e$ component that cannot be depreciable (see red box), because it is of the same order of magnitude of $e^2/C_q$, resulting in a degeneracy of energy $(g_r e^2/C_q)$ for the ET reaction (see discussion conducted in section~\ref{sec:degeneracy}). This $e^2/C_e$ contribution to $e^2/C_q$ is governed by a length scale of $l$, which is much lower than that of $L_{nf}$.}
\label{fig:k-charge-modes}
\end{figure}

The latter analysis of charging an electrochemical interface in the absence of an electrochemical reaction, as studied by Goes et al. ~\citep{Goes-2012, Lehr-2017}, is depicted in Figure~\ref{fig:k-charge-modes} as a series resistive-capacitive circuit, i.e., $R_{nf}C_\mu$, where $1/C_q$ is absent in $1/C_\mu = 1/C_e + 1/C_q$, resulting in $R_{nf}C_e$. Therefore, non-faradaic charging time-dependent dynamics will apply. These dynamics are governed by a non-faradaic $k_{nf} = G_{nf}/C_e$ rate, where $ G_{nf} = 1/R_{nf}$ has an associated non-faradaic current of $i_{nf} = C_e s$, in which $s = dV/dt$ is the scan rate or a potential-time dependent perturbation that access the charge dynamics associated with $C_e$. 

Specifically, for this situation, the non-faradaic conductance is defined as $G_{nf} = 1/R_{nf}$ and $C_e$, as illustrated in Figure~\ref{fig:k-charge-modes}, is a geometric capacitance per unit of area of the electrode. The magnitude of $C_e$ is given by $C_e = \epsilon \delta_D$, where $\epsilon = \epsilon_r \epsilon_0$. $\epsilon$ is the dielectric constant of the environment, where $\epsilon_r$ is the relative dielectric constant and $\epsilon_0$ the vacuum electric permittivity constant. Note that $\delta$ is the reciprocal of the Debye length, i.e. $\delta_D \propto 1/L_{nf}$, where the Debye length is proportional to $L_{nf}$, the charge separation between the electrode and the anions or cations in the solution/electrolyte interface adjacent to the monolayer (see Figure~\ref{fig:k-charge-modes}). This $L_{nf}$ is dependent on the concentration of the ions in the bulk of the solvent, as required from a thermodynamics description and modeling of a double-layer capacitance~\citep{Bard-book, Schmickler-book} through a Debye length assumption.

\begin{figure}[h]
\centering
\includegraphics[width=7cm]{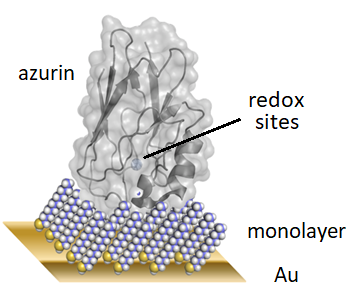}
\caption{Redox film comprised of a non-electro-active alkanethiol monolayer covalently attached to a gold electrode. Redox proteins (Azurin) can be attached to the alkanethiol monolayer, and an additional protein film can be formed over the monolayer. Reprinted (adapted) with permission from Paulo R. Bueno, Giulia Mizzon, Jason J. Davis; Capacitance Spectroscopy: A versatile approach to resolving the redox density of states and kinetics in redox-active self-assembly monolayers, The Journal of Physical
Chemistry B. Copyright (2012) American Chemical Society.}
\label{fig:Azurin-film}
\end{figure} 

\begin{figure*}[h]
\centering
\includegraphics[width=17.5cm]{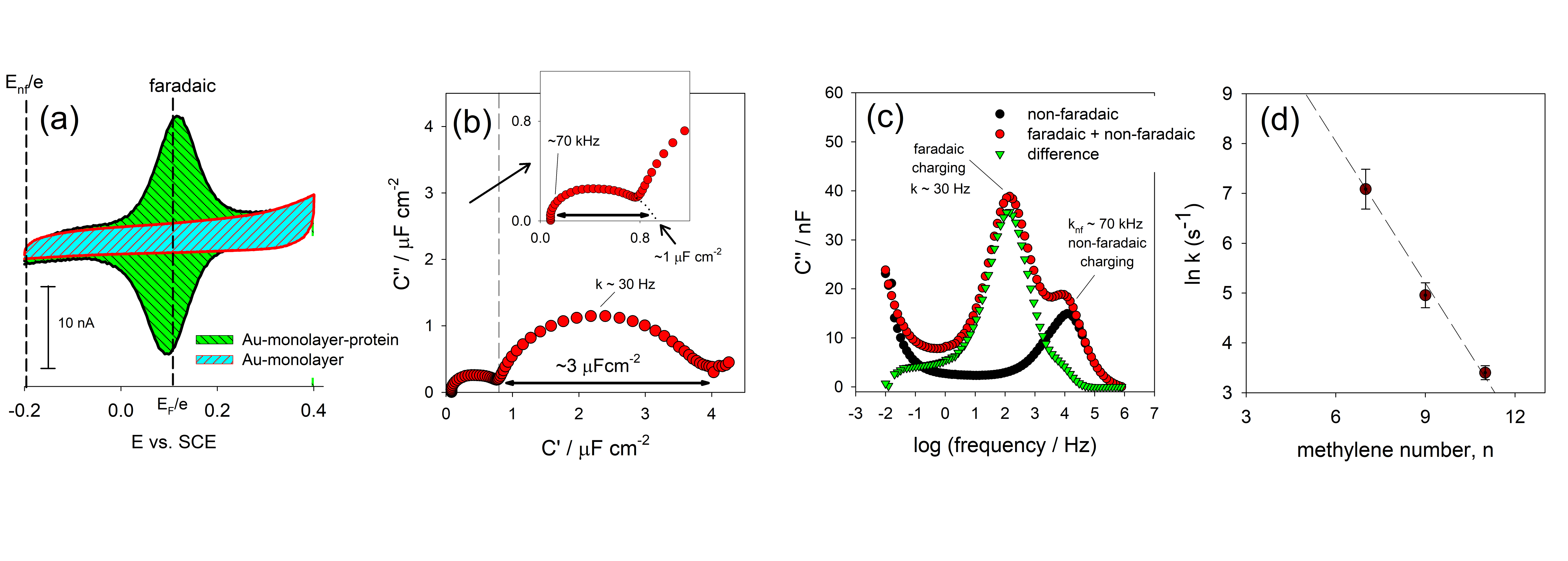}
\caption{(a) Cyclic voltammogram of a film comprising a non-electro-active monolayer self-assembled on gold within Azurin-redox-active proteins added on the top of this monolayer, as depicted in Figure~\ref{fig:Azurin-film}. In green, the redox electrochemical current contribution is shown, and cyan is the isolated response of the monolayer without Azurin on it. (b) A complex capacitive spectrum (Nyquist diagram) of the response of the film measured at the formal potential $E_F/e$, as denoted in (a), where two modes of charging the film are evidenced. (c) The same scenario as in (b), but the imaginary component of the complex capacitance is depicted as a function of frequency, where the peaks correspond to the resonance frequencies associated with non-faradaic $k_{nf}$ and faradaic $k$ charging modes, as elaborated on in the text. (d) Linear dependence of the logarithmic of the faradaic charging mode $k$ as a function of different lengths $L$ of the non-active monolayer. The linear dependence confirms that the electron coupling occurs by tunneling, i.e. $\sum_{n=1}^{N}T_{n}\left( \mu \right) = \exp \left( -\beta L \right)$. Reprinted (adapted) with permission from Paulo R. Bueno, Giulia Mizzon, Jason J. Davis; Capacitance Spectroscopy: A versatile approach to resolving the redox density of states and kinetics in redox-active self-assembly monolayers, The Journal of Physical
Chemistry B. Copyright (2012) American Chemical Society.}
\label{fig:Azurin-film-mesur}
\end{figure*}

From an experimental perspective, consider the measurement of the capacitance of an interface containing a non-electro-active alkanethiol monolayer~\citep{Goes-2012, Lehr-2017}, which is covalently attached to an electrode. A redox-protein (azurin) was attached to the monolayer, as illustrated in Figure~\ref{fig:Azurin-film}. The current-voltage response (voltammogram) of the film is shown in Figure~\ref{fig:Azurin-film-mesur}\textit{a}, where the electrochemical current response of the monolayer solely is shown in cyan color, and the response within the Azurin redox-protein over it is shown in green.

The redox peaks in the voltammogram are attributed to the redox activity of the protein. Figure~\ref{fig:Azurin-film-mesur}\textit{b} shows the complex capacitance spectra, which are in agreement with the dynamics described by Eq.~\ref{eq:Complex-Cq}. The complex capacitive spectrum, which follows Eq.~\ref{eq:Complex-Cq}, was measured in the formal potential $E_F/e$, as indicated in Figure~\ref{fig:Azurin-film-mesur}\textit{a}. Two parallel RC charging processes can be observed, each of which can be interpreted as a particular charging mode of the interface, in compliance with Eq.~\ref{eq:k-Celect-zeroT}. The total rate response is composed of two contributions $k_{nf} + k$, each of them obeying a different RC dynamics. In other words, each of these rate modes has a different physical meaning of the RC dynamics, representing different charging mechanisms and modes of time-dependent responses of the electrochemical interface. 

In other words, one of these rates is attributed to a non-faradaic charging mode $k_{nf} = G_{nf}/C_e$ rate, as was described above, whereas the other mode is related to a faradaic $k = G/C_q$ type rate\footnote{Particularly, this faradaic charging mode follows Eq.~\ref{eq:Complex-Cq}, in with $k = 1/\tau = 1/R_qC_q$. $C_q^0$ is the equilibrium capacitance that governs the equilibrium charge state of the interface.}. Note that $G_0\sum_{n=1}^{N}T_{n}\left( \mu \right)$ are different for each of these charging modes of the interface because they possess different electron scattering mechanisms governed by different $\sum_{n=1}^{N}T_{n}\left( \mu \right)$ probabilities. The higher frequency ($\sim$ 70 kHz) observed experimentally is attributed to the non-faradaic charging process $k_{nf} = G_{nf}/C_e = 1/R_{nf} C_e$ with a $C_e$ of $\sim$ 0.8 $\mu$F cm$^{-2}$, whereas the lower frequency ($\sim$ 30 Hz) charging mode is a $k = G/C_q = 1/R_qC_q$ rate with a capacitance of $\sim$ 3 $\mu$F cm$^{-2}$.

The Bode diagram, in Figure~\ref{fig:Azurin-film-mesur}\textit{c}, shows the imaginary capacitive component of the complex capacitive spectrum as a function of the perturbation frequency. This figure suitably shows the two charging dynamic modes of the interface, where the non-faradaic response is measured at a potential of -200 mV (indicated by a black vertical dashed non-faradaic line in Figure~\ref{fig:Azurin-film-mesur}\textit{a}, as $E_{nf}/e$) and faradaic-response is measured at the formal potential of $E_F/e$ (indicated by a black vertical dashed faradaic line in Figure~\ref{fig:Azurin-film-mesur}\textit{a}). Note the superposition of both non-faradaic and faradaic responses. 

The non-faradaic rate dynamics can be subtracted from the total rate response, a spectroscopic procedure that is shown in green in Figure~\ref{fig:Azurin-film-mesur}\textit{c}. This subtracted faradaic rate spectroscopic method demonstrates that by employing impedance-derived capacitance spectroscopy~\citep{Bueno-2012}, the relativistic quantum electrodynamics associated with the quantum rate, as defined in Eq.~\ref{eq:nu}, can be studied in detail. 

For instance, the electron coupling associated with the $k$ mode of charging the interface, which follows a tunnelling regime of electron transport according to the quantum rate theory applied to redox reactions, is confirmed in Figure~\ref{fig:Azurin-film-mesur}\textit{d}, where $\sum_{n=1}^{N}T_{n}\left( \mu \right) = \exp \left(-\beta L \right)$ is obeyed. The tunneling regime of the ET dynamics is confirmed by plotting the logarithmic of $k$ (faradaic) against the thickness $L$ of the alkenathiol monolayer, which, according to the quantum rate theory, represents the length of the quantum channels.

\begin{figure}[h]
\centering
\includegraphics[width=8cm]{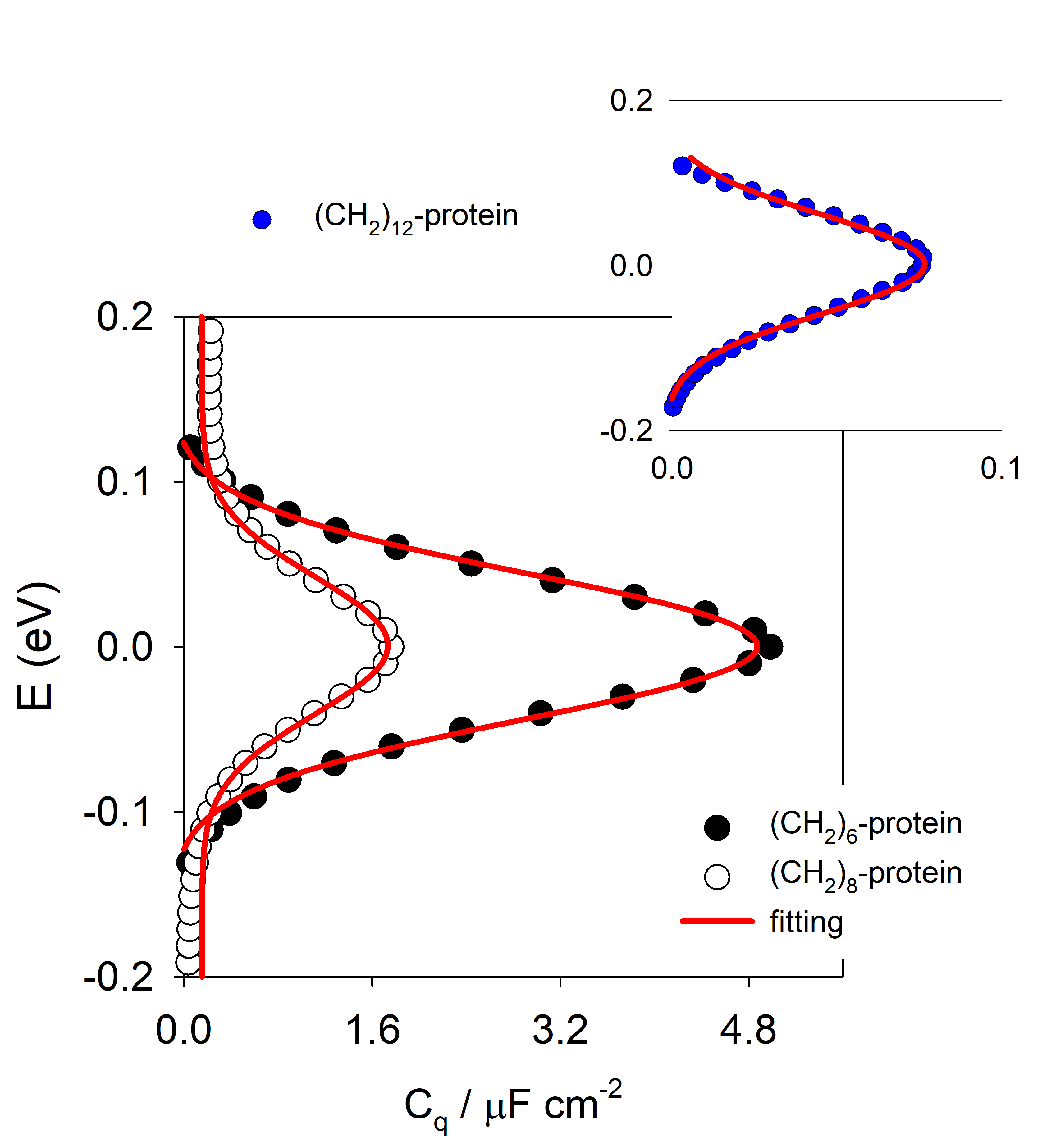}
\caption{Electrochemical density-of-states (EDOS) shape inferred from $C_q = e^2 \left( dn/dE \right)$ and measured for three different electronic coupling modes (three different thicknesses $L$, where the inset is the response for the (CH)$_{12}$-protein film) of the redox-active protein film to the electrode, where the access to the DOS is restricted owing to the thickness of the non-electro-active alkanethiol layers (CH$_i$, where $i$ = 6, 8 and 12), lowering the electronic coupling of the redox states to the electrode, as it is governed by $\sum_{n=1}^{N}T_{n}\left( \mu \right) = \exp \left( -\beta L \right)$. Reprinted (adapted) with permission from Paulo R. Bueno, Giulia Mizzon, Jason J. Davis; Capacitance Spectroscopy: A versatile approach to resolving the redox density of states and kinetics in redox-active self-assembly monolayers, The Journal of Physical Chemistry B. Copyright (2012) American Chemical Society.}
\label{fig:DOS-azurin}
\end{figure}

Note that the DOS $\left( dn/dE \right) = C_q/e^2$ of Azurin redox films can be obtained by measuring $C_q$ for $\omega \rightarrow 0$, which in Eq.~\ref{eq:Complex-Cq} corresponds to the magnitude of the value of $C_q^0$. Experimentally, this is obtained by measuring $C_q$ at the frequency where the second semi-circle of Figure~\ref{fig:Azurin-film-mesur}\textit{b} closes. Once this frequency is identified, a potential scan is conducted and $C_q^0$ is measured as a function of different energy potential levels of the electrode, as shown in Figure~\ref{fig:DOS-azurin}. In Figure~\ref{fig:DOS-azurin}, the redox DOS, corresponding to $C_q^0$ as a function of electrode energy level, is depicted for three different redox-inactive monolayer lengths. Note that the accessibility to the redox DOS of Azurin films varies according to $\sum_{n=1}^{N}T_{n}\left( \mu \right) = \exp \left(-\beta L \right)$, in agreement with Figure~\ref{fig:Azurin-film-mesur}\textit{d}. 

As expected, the electronic coupling of the redox states to the electrode was lower for longer redox-inactive monolayer methylene lenghts. The correctness of the analysis and interpretation of the quantum rate theory for these electrochemical reactions is evident. This is based not only on the analysis of the electronic coupling of the states to the electrode, but also in the analysis of the fitting of these redox DOS experimental curves to Eq.~\ref{eq:Cq-thermal}, as shown by the red line curves in Figure~\ref{fig:DOS-azurin}, revealing that the theoretical and experimental results are in good agreement.

Before concluding this section, it is important to describe the equivalent circuit of an electrochemical interface comprising of dielectric polarization (of the thiolated bridge) and quantum faradaic (of the azurin moieties) charging modes, as theoretically and experimentally studied in this section. The films studied in this section were comprised of a gold electrode (as an electron reservoir) modified with a non-electro-active thiolated film, over which another layer of azurin was assembled as a redox-active protein layer (see Figure~\ref{fig:Azurin-film}).

\begin{figure}[h]
\centering
\includegraphics[width=5cm]{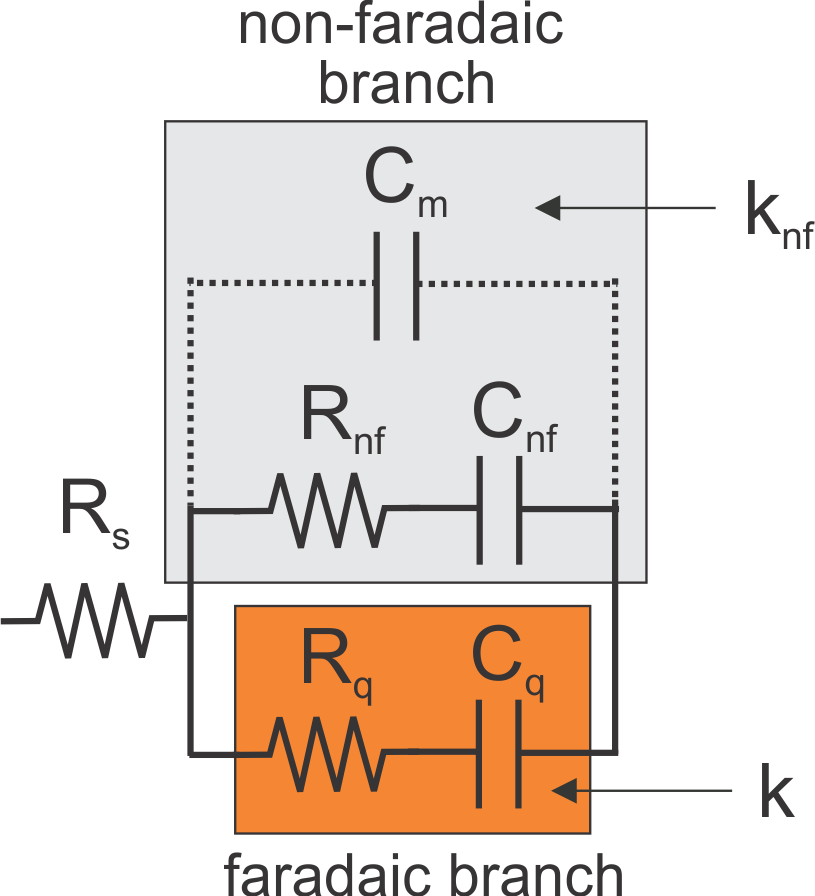}
\caption{Equivalent circuit of an electrochemical interface comprising of non-faradaic $k_{nf} = 1/R_{nf}C_{nf}$ and faradaic $k = 1/R_qC_q$ charging modes, as discussed in the text. $R_s$ is the series resistance comprising the contact $R_c$ and electrolyte $R_e$ resistances. $C_m$ is the higher frequency capacitive response of the interface (see that in Figure~\ref{fig:Azurin-film-mesur} this corresponds to a value lower than 0.1 $\mu$F cm$^{-2}$ in the beginning of the first semi-circle). $C_m$ tends to be much lower than $C_{nf}$ and albeit it can be ignored in a fitting of the experimental data (without affect a quantitative analysis of the results), its accountancy is important to conform with Debye or Cole-Cole dielectric relaxation, which suitably describe this type of non-faradaic dielectric dipolar relaxation~\citep{Goes-2012, Lehr-2017}.}
\label{fig:circuit-azurin}
\end{figure}

The theoretical equivalent circuit that fits the experimental data is shown in Figure~\ref{fig:circuit-azurin}, where two parallel charging modes ($k_{nf}$ and $k$) are evidenced, both of which are in agreement with the general rate description provided by Eq.~\ref{eq:k-Celect-zeroT}.  Note that $C_m$ is the higher frequency response of the capacitance of the monolayer and $R_s$ is the series resistance comprising  the contact $R_c$ and electrolyte $R_e$ resistances. The circuit depicted in gray is the classical Debye or Cole-Cole dielectric relaxation which is discussed in detail in references~\cite {Goes-2012, Lehr-2017}. The circuit described in red follows the relativistic quantum dynamics predicted by Eq.~\ref{eq:Complex-Cq} and with the physical meaning provided by the quantum rate theory, where $k = 1/R_qC_q = 1/\tau$ is in agreement with the equivalent circuit studied in Figure~\ref{fig:QRC-circuit}.

The role of the electrolyte (including counter-ions electric field screening with a length $l$) over $C_q$ quantum states is a particular assumption to be discussed further that can be interpreted as an additional energy degeneracy. Therefore, observe that there is a series $C_e$ capacitive component associated with $C_q$, as noted in red in Figure~\ref{fig:k-charge-modes}. This $C_e$ is governed by a length scale of $l$ that accounts for the action of the electrolyte and ions over the quantum states within Ox/Red moieties of the redox monolayer, which can be considered an electric energy degeneracy in $e^2/C_q$ besides that associated with $g_s$. This is stated as a consequence that $C_e$ can be, in the presence of an appropriate electrolyte electric-field screening, of the same magnitude of $C_q$, as will be discussed in-depth in section~\ref{sec:degeneracy}.

In the next section, the meaning of the electrochemical (redox) density-of-states (EDOS) within the quantum rate theory will be discussed.

\section{The Meaning of the Electrochemical (Redox) Density-of-States}\label{sec:Meaning-DOS}

In the present section, the characteristics and meaning of EDOS observed in redox monolayer films will be discussed. We will focus only on the faradaic $k$ models of charging redox films. This type of EDOS is of great interest because it represents a particular electrochemical reaction that can be well-controlled and served as an experimental prototype for studying quantum electrochemistry and relativistic ET dynamics within the quantum rate theory.

\begin{figure*}[h]
\centering
\includegraphics[width=12cm]{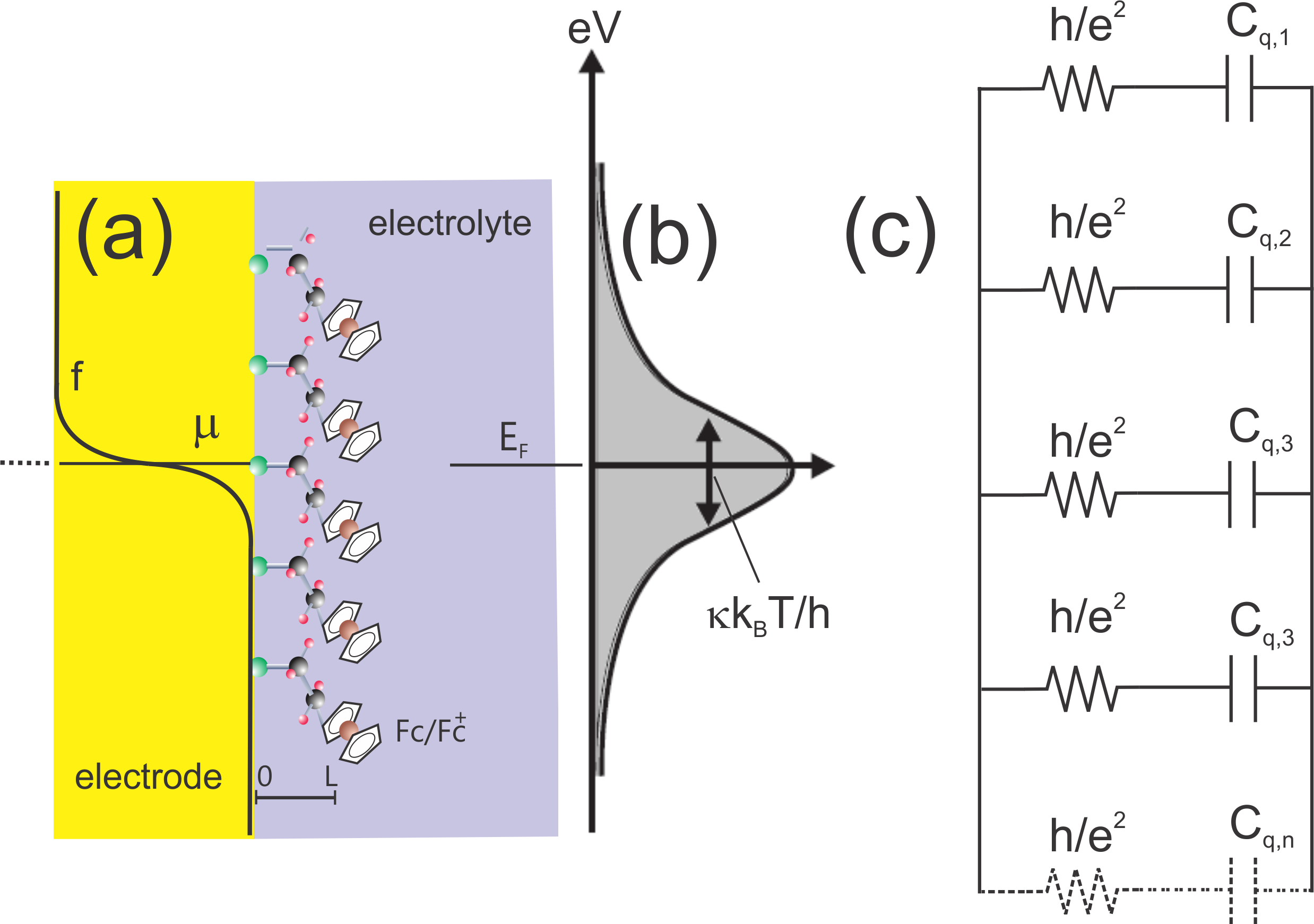}
\caption{Schematic depiction of a redox molecular film assembled over a gold electrode. (a) Thiolated ferrocene monolayers self-assembled over a gold electrode. (b) Maximum of the redox activity and the value of the redox or quantum capacitance of the interface is obtained when the electrochemical potential of the electrode $\mu$ is poised at the same level of the formal potential $E_F/e$ of the monolayer. As the shape of $C_q$ controls the properties of the interface, in agreement with Eq.~\ref{eq:Cq-thermal}, the pre-exponential term $\kappa k_BT/h$ of Eq.~\ref{eq:Marcus-quantum} is related to the broadening of the DOS or the shape of $k$ as a function of the electrode potential level. (c) Depicts the quantum RC statistical mechanical ensemble where each molecule in the electrode has an individual quantum RC dynamics in agreement with the quantum rate theory that predicts a DOS shape in agreement with Eq.~\ref{eq:Cq-thermal} and Eq.~\ref{eq:Marcus-quantum}.}
\label{fig:Fc-film-model}
\end{figure*}

In particular, we focused on the meaning of the EDOS obtained in a ferrocene-thiolated self-assembled monolayer (Fc-SAM), as schematically depicted in Figure~\ref{fig:Fc-film-model}\textit{a}. This Fc-SAM film serves as a model of different redox-tagged monolayers because, equivalent to the azurin film studied in the previous section, the phenomenology is equivalent to any molecular redox films, which have been useful for studying the heterogeneous ET dynamics of redox reactions~\citep{Sanchez-2022-1, Alarcon-2021}. The absence of a diffusional kinetic limitation permits the study of quantum mechanical properties of these redox reactions at room-temperature~\citep{Bueno-book-2018}. 

Although the direct contact of a metallic electrode with redox species chemically attached to the electrode allows us studying the ET reaction, the use of free redox species in a solution/electrolyte phase in direct contact with a metal is a type of ET reaction that has classical mass-diffusional kinetic control; hence, it is not suitable for studying the quantum mechanics of ET reactions. On the other hand, the electrochemistry of adsorbed or covalently attached redox-active moieties on metallic electrode permits the establishment of the quantum mechanical meaning of key experimental parameters of the interface, such as exchange current $i_0$, charge-transfer resistance $R_{ct}$, $C_q$ and $k$. For more details, see reference~\citep{Sanchez-2022-2} and sections \ref{sec:QRct} and \ref{sec:QR-mediation}. 

The diffusionless characteristic of this type of ET process allows a suitable experimental setting that permits direct access to the EDOS through the measurement of $C_q$. Owing to the experimental accessibility of the EDOS, this type of interface can be used as a model to study quantum electrochemistry using different experimental settings, which allows us to study the interfacial electrochemical reactions at the molecular scale with excelent precision and resolution.

\begin{figure*}[h]
\centering
\includegraphics[width=18cm]{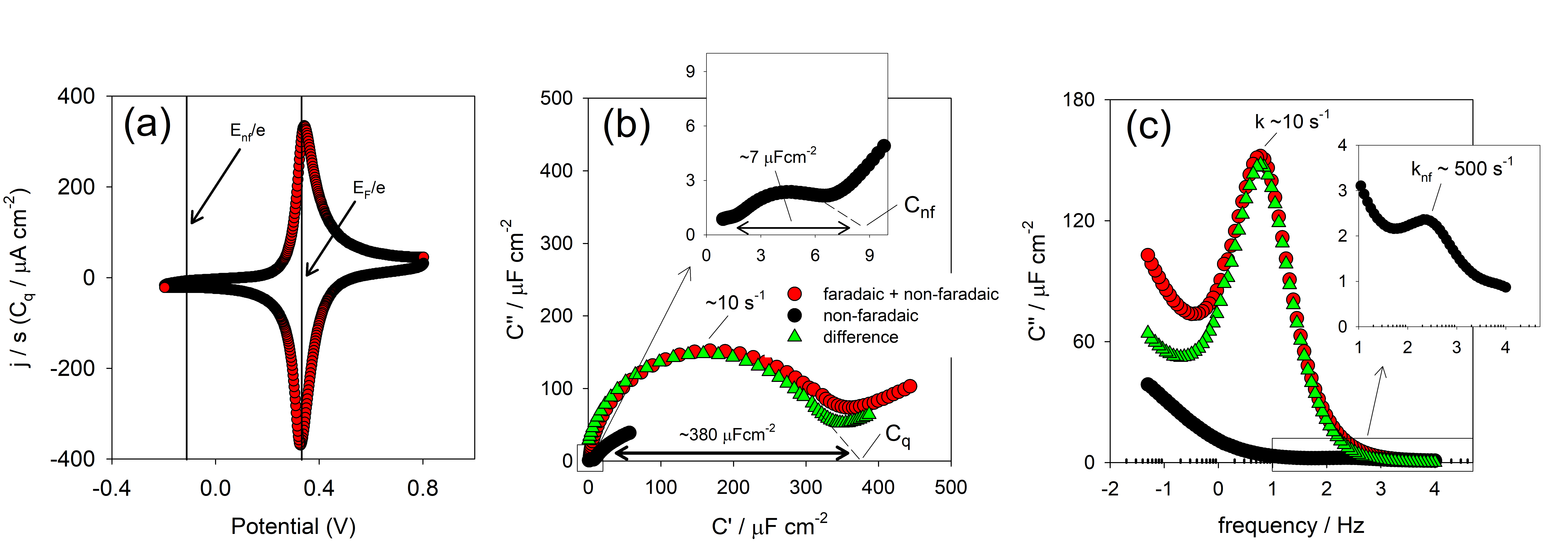}
\caption{(a) Capacitance-voltage curve (as an alternative depiction of cyclic voltammograms) showing the dominance of the redox activity of a thiolated ferrocene monolayer over the non-faradaic mode of charging the interface. $E_{nf}/e$ is a referential electrode potential where impedance-derived capacitive spectra are obtained to investigate the contribution of the non-faradaic charging mode of the interface. $E_{F}/e$ corresponds to the formal potential, where the faradaic activity is maximum. (b) Nyquist and (c) imaginary Bode capacitive diagrams of non-faradaic and faradaic responses of the interface. As was the case discussed for Azurin film (Figure~\ref{fig:Azurin-film-mesur}), the parallel non-faradaic mode of charging the interface can be subtracted from the total. The remaining response is predominantly related to the faradaic activity. In both (b) and (c) diagrams the dominance of the faradaic activity over the non-faradaic is apparent. This allows us to use these types of redox monolayers as a suitable model to study quantum electrochemistry using the equivalent circuit model depicted in Figure~\ref{fig:QRC-circuit}, whose ET reaction dynamics that are in agreement with the relativistic quantum rate dynamics. Reprinted (adapted) with permission from Paulo R. Bueno and Jason J. Davis; Elucidating Redox-Level Dispersion and Local Dielectric Effects within Electroactive Molecular Films, Analytical Chemistry. Copyright (2014) American Chemical Society.}
\label{fig:Fc-film-experimental}
\end{figure*}

Similarly to the case of the Azurin films studied in the previous section, in a redox Fc-SAM there are also two modes of charging the interface: non-faradaic and faradaic~\citep{Bueno-2014-2}. However, for the azurin films, these two charging contributions were similar in magnitude, which was not the case for the Fc-SAM films. Despite the different magnitudes of the charging modes in the Fc-SAM films, the charging of the interface is equivalently discussed in terms of two RC-parallel contributions, which is in agreement with the equivalent circuit model depicted in Figure~\ref{fig:circuit-azurin}. In summary, the same phenomenological charging mode analysis applies to the Fc-SAM films, as shown in Figure~\ref{fig:Fc-film-experimental}, and the only differences between the Azurin and Fc-SAM films is in the magnitude of the circuit elements.

Figure~\ref{fig:Fc-film-experimental}\textit{a} shows the voltammetric responses of the Fc-SAM films. In this figure, the Faradaic (redox) response is more dominant than the non-faradaic response. The voltammogram is represented as the electrochemical current density normalized by the scan rate $s$, which permits the direct measurement of $C_q$ of the interface. The dominance of $C_q$ is demonstrated based on the magnitude of the capacitances. The non-faradaic response is a $C_{nf}$ capacitance of $\sim$ 7 $\mu$F cm$^{-2}$ (shown in the inset of Figure~\ref{fig:Fc-film-experimental}\textit{b}), whereas the magnitude of the $C_q$ faradaic response is of $\sim$ 380 $\mu$F cm$^{-2}$. 

These differences in non-faradaic $C_{nf}$ and faradaic $C_q$ capacitances indicate that 98\% of the capacitive response of the interface is due to $C_q$, as confirmed by Figure~\ref{fig:Fc-film-experimental}\textit{c}. In this figure, it can be observed that, whether the non-faradaic capacitive is subtracted from the total response, there is a minimal difference in the faradaic relaxation peak before and after the subtraction, which demonstrates that the influence of the non-faradaic response on the faradaic response is minimal. The frequency of the non-faradaic charging mode is observed in the inset of Figure~\ref{fig:Fc-film-experimental}\textit{c}, which also demonstrates that the rate of each of these modes of charging the interface is also significantly different.

The dominance of the faradaic response over the non-faradaic response of the for Fc-SAM films, which has a different magnitude than that observed in azurin films, is explained by the higher redox DOS of the interface. This higher DOS is proportional to the number of redox states per unit area, referred to as molecular coverage, which is higher in the Fc-SAM than in Azurin films. The number of these redox states corresponds to the quantum rate theory of $N$ quantum modes or quantum channels. In Azurin films, the redox activity is computed by a $C_q$ of barely $\sim$ 3 $\mu$F cm$^{-2}$, whereas in Fc-SAM it corresponds to a $C_q$ of $\sim$ 380 $\mu$F cm$^{-2}$. Whether the molecular material covalently attached to the interface is normalized per gram of molecular material, this value of $\sim$ 380 $\mu$F corresponds to an astonishing specific capacitance of $\sim$ 2,000 F/g for the the interface. 

Accordingly, Fc-SAM films have remarkable pseudo-capacitive characteristics and a super-capacitance performance that has not been achieved in supercapacitor and battery devices comprising electro-active electrodes assembled on a microscopic scale.
The nanoscale origin of the pseudo-capacitance~\citep{Bueno-2019} is associated with redox dynamics at the atomic scale, which have an intrinsic origin as a manifestation of a quantum mechanical phenomenon within a quantum RC electrochemical dynamics, meaning that this RC dynamics cannot be understood without considering the relativistic characteristics of the dynamics coupled to the electrolyte environment. Pseudo-capacitive phenomena observed in supercapacitors and battery devices are the macroscopic manifestation of this quantum ET phenomenon. For instance, in the case of Prussian Blue and analogous compounds, the intrinsic observable pseudo-capacitance is associated with the exchange electrodynamics~\citep{Bueno-2006} between Fe$^{3+}$(NC)$_6$ and Fe$^{2+}$(NC)$_6$ sites within these compounds. Equivalent to the case of redox monolayers, in which the measurable pseudo-capacitance can be applied for sensing biomarkers of interest~\citep{Garrote-2020}, the pseudo-capacitance of Prussian Blue compounds can be used for sensing molecules~\citep{Oliveira-2019}, demonstrating that the concept of electrochemical capacitance described in this manuscript can be extended further and is not restricted to redox monolayers.

Hence, the study of the meaning of the DOS in Fc-SAM films allows us to comprehend the origin of the pseudo-capacitive phenomenon in great detail, which contrasts with the classical analysis. The pseudocapacitive phenomenon studied from a quantum mechanical approach implies some level of contradiction with the common classical sense of the correlation between capacitance and conductance phenomena. Classical electrostatic capacitance is based on charge-storage mechanisms that is governed by dielectric polarization (within a spatial separation of charge), which implies the maximum capacitance for minimal conductance. Nonetheless, in contrast to this classical understanding, pseudo-capacitive phenomena, which follows relativistic ET dynamics, imply that the maximum capacitance is accompanied by a maximum conductance.

\begin{figure}[h]
\centering
\includegraphics[width=8cm]{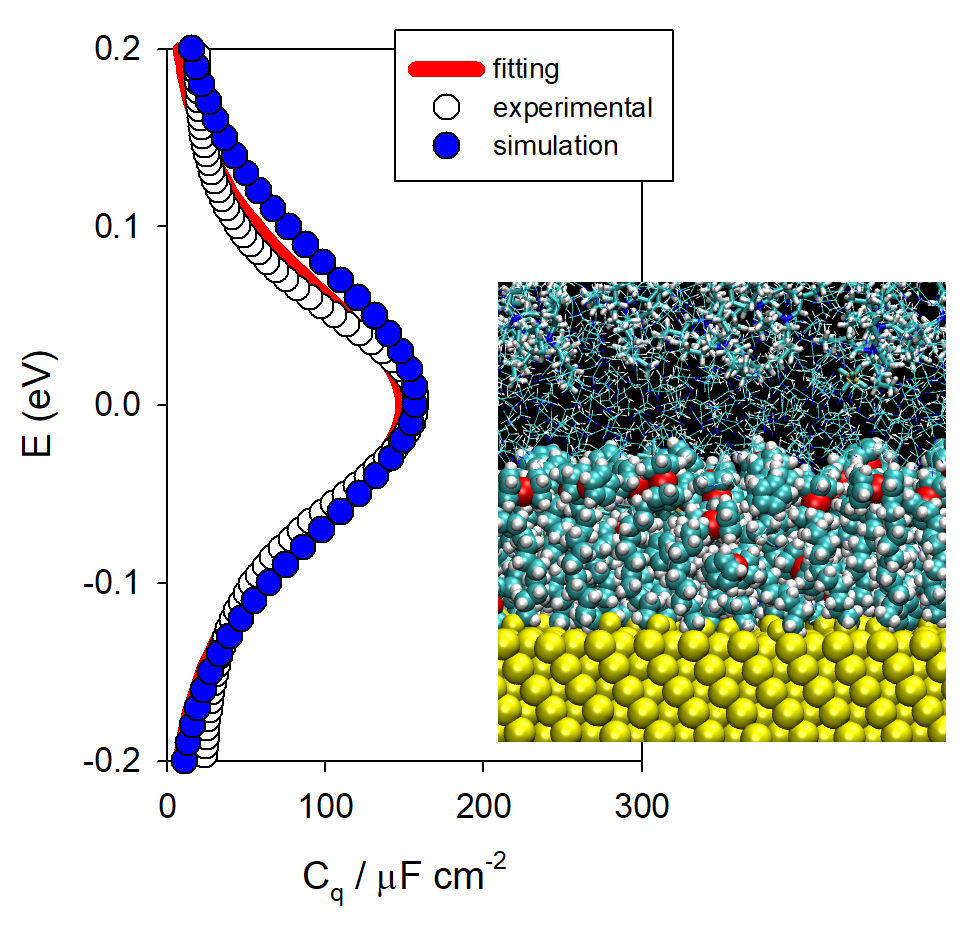}
\caption{Electrochemical DOS for Fc-SAM films depicted as a function of $C_q = e^2 \left( dn/dE \right)$. The uncolored dots depict the experimental data obtained by the impedance-derived capacitive method and the red line corresponds to the fitting of the experimental data to Eq.~\ref{eq:Cq-thermal}. In blue dots is shown the result obtained from computational simulation methods that consider the potential contribution of the gold electrode as well as of the solvent environment, as schematically depicted in the inset.}
\label{fig:Fc-CPmethod}
\end{figure}

For instance, in quantum rate theory each molecule attached to a gold electrode behaves as a quantum RC circuit, as illustrated in Figure~\ref{fig:Fc-film-model}\textit{c}. A statistical mechanical ensemble of these quantum conductive $e^2/h$ and capacitive $C_q$ elements leads to a total $C_q$ capacitance that is calculated using Eq.~\ref{eq:Cq-thermal}, with a theoretical response depicted in Figure~\ref{fig:Fc-film-model}\textit{b}. A detailed computational simulation~\citep{Bueno-2021} conducted in Fc-SAM films demonstrates the agreement between the quantum rate theory, experimental EDOS, and EDOS simulated by computational methods, as shown in Figure~\ref{fig:Fc-CPmethod}. The good agreement between the quantum rate theory, experiments, and simulation is an additional validation of the quantum rate principle that enhances our understanding of the ET reaction dynamics.

In the next section, the molecular and electronic structures of the redox self-assembly monolayers will be discussed. As redox self-assembly monolayers constitute a particular experimental molecular model for studying quantum electrochemistry, it is important to discuss the molecular and electronic structures of the films in detail. This knowledge was obtained from studies on the EDOS using computational calculations, which revealed important characteristics of the EDOS.

\section{Molecular and Electronic Structures of Redox Monolayers}\label{sec:StrElec-EDOS}

\begin{figure}[h]
\centering
\includegraphics[width=8cm]{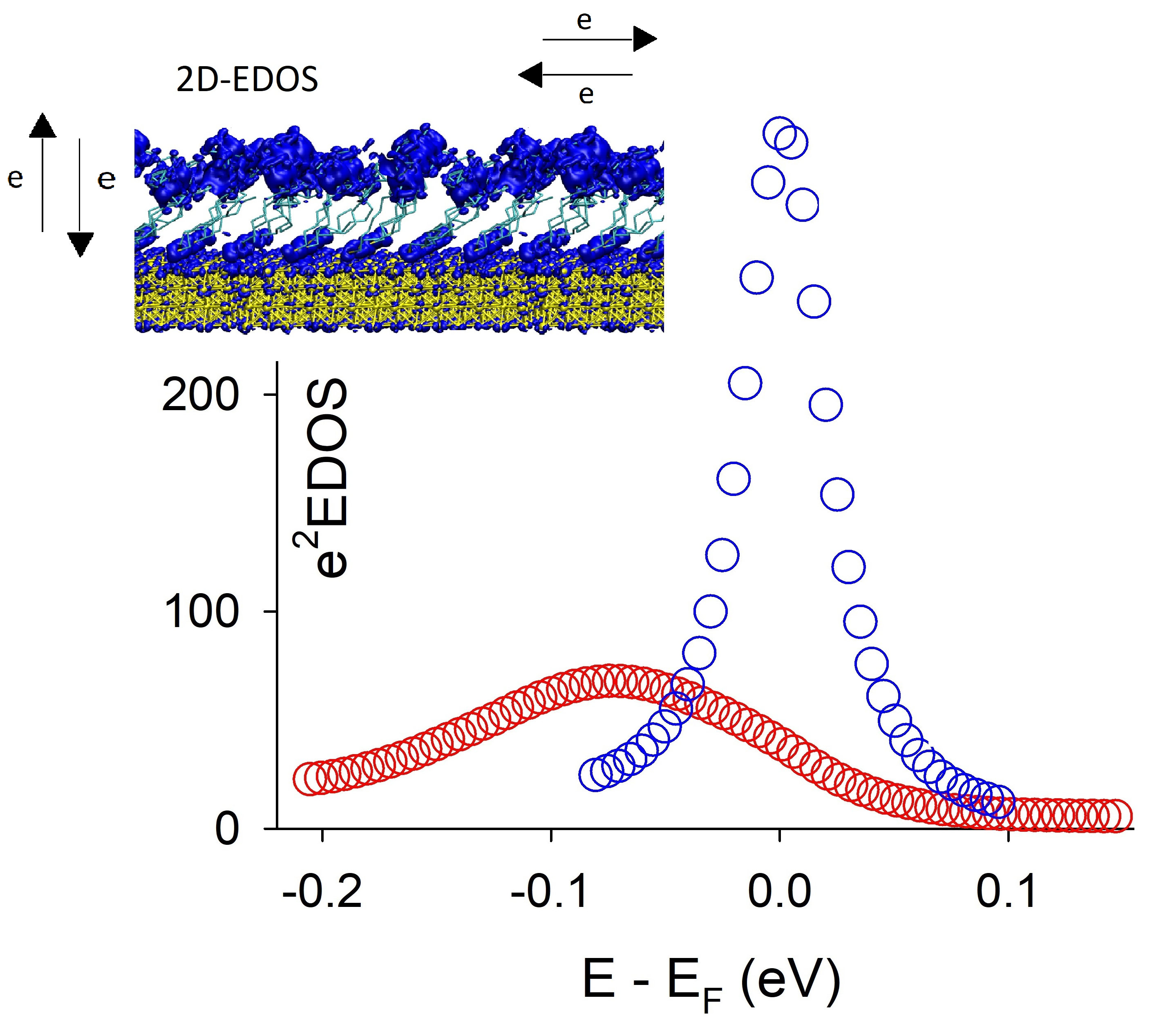}
\caption{Molecular and electronic structure calculations (inset) of Fc-SAM films demonstrating that the EDOS has a 2D-like structure where most of electron density is located close to the Fc$^{+}$/Fc ET dynamics (indicated in Figure~\ref{fig:DA-electrode-electrolyte}) occurring at a distance of $\sim$ 2 nm from the electrode. This intrinsic Ox + e $\rightleftharpoons$ Red electrochemical dynamics is different from that occurring between Fc$^{+}$/Fc states and the electrode. The above description of the EDOS dynamics, that is, with a dynamics perpendicular and parallel to the plane of the electrode, is depicted in the inset and was observed from the calculation of the electronic structure obtained at the formal potential state of the working electrode. The red curve is the experimental EDOS obtained for a Fc-SAM in an aqueous solvent (dielectric constant of $\sim$ 80), and the blue curve is the EDOS of the same Fc-SAM film in acetonitrile (dielectric constant of $\sim$ 40). The number of states (about $10^{12}$) of these EDOS curves (calculated from the integral of the curve) is invariant with changes in the dielectric constant of the environment, meaning that the role of the environment, as an external potential acting over $D$ and $A$ quantum states, only has the effect of spreading the redox energy levels. Reprinted (adapted) with permission from Gustavo T. Feliciano and Paulo R. Bueno; Two-Dimensional Nature and the Meaning of the Density of States in Redox Monolayers, The Journal of Physical Chemistry C. Copyright (2020) American Chemical Society.}
\label{fig:Fc-QMMM}
\end{figure}

Evidence for the ET dynamics parallel to the plane of the electrode, as depicted in Figure~\ref{fig:DA-electrode-electrolyte}, was confirmed using quantum mechanical and molecular dynamical hybrid computational simulation methods~\citep{Feliciano-2020}, which enabled the the electronic structure of the redox monolayer to be calculated. Hence, it was confirmed that redox dynamics existed between the Ox/Red moieties within the regular redox film, as shown in the inset of Figure~\ref{fig:Fc-QMMM}. This additional internal ET dynamics, also schematically depicted in Figure~\ref{fig:DA-electrode-electrolyte}, confirmed that a higher EDOS exists at the formal potential $E_F/e$ and indicated that redox Fc-SAM films assembled over conductive electrodes resemble, in terms of the electronic structure of the EDOS, two-dimensional electron gas (2DEG) structures, as introduced by Serge Luryi~\citep{Luryi-1988}.

According to Luryi~\citep{Luryi-1988}, a quantum capacitor device can be designed if one of the plates of a planar capacitor is composed of a very thin conductive layer separated from a continuum planar macroscopic electrode by a molecular dielectric bridge with a thickness of $<$ 2 nm. This Luryi's definition of a quantum capacitor device agrees with the general meaning of quantum capacitances, as expressed in Eq.~\ref{eq:Cq-general}. This definition of quantum capacitance predicts that if one of the plates is a planar capacitor device with a relatively small DOS (e.g., a molecular film with quantum redox states) and the other has a larger DOS (e.g., a metallic plate), the structure behaves like a quantum capacitor, in which one of the plates acts as a macroscopic lead. 

This depiction of a quantum capacitor using a planar configuration resembling classical plate capacitors was originally outlined by Luryi~\citep{Luryi-1988} as a possible macroscopic method of designing quantum capacitors. Hence, the EDOS of the monolayers is a type of quantum capacitive device structure that follows the architecture proposed by Luryi~\citep{Luryi-1988} for constructing solid-state quantum capacitors.

In other words, well-organized Fc-SAM besides serving as an appropriate model for studying ET reaction in heterogeneous diffusionless settings, it is intrinsically a quantum molecular devices that permits the study of quantum electrodynamics in an electrochemical/electrolyte environment settings. This not only provides additional ways of studying pseudocapacitive phenomena~\citep{Bueno-2019}, but also permits the comparison of nanoscale electronics and electrochemistry~\citep{Bueno-2018}. Molecular dynamics calculations combined with quantum mechanics computational methods also confirmed the rigidity of the redox monolayer, in which redox moieties are placed within a well-defined distance $L$ of the electrode~\citep{Feliciano-2020}.

Additionally, computational simulation methods confirmed that the ET dynamics between the electrode and the EDOS plane is different of Ox/Red ET dynamics, i.e. Ox + e $\rightleftharpoons$ Red, that occurs internally and in parallel to the plane of Fc-SAM film, as shown in the inset of Figure~\ref{fig:Fc-CPmethod}. This internal Ox/Red ET dynamics is the origin of the 2DEG-like structure thin conductive layer suggested by Luryi~\citep{Luryi-1988}. The separation of this planar Ox/Red ET conductive dynamics from the electrode plate has a length $L$ $\sim$ 1.8 nm separation from the electrode within a dielectric polar structure acting as a non-electric conductive molecular bridge. The length $L$ of the separation of the redox conductive layer from the electrode corresponds to the length of the quantum channels, which permits a measurable redox capacitive dynamics. Hence, $L$ is the length that governs the redox ET dynamics with the electrode within a quantum rate dynamics of $k$.

In the next section, we demonstrate that if well-behaved redox-active monolayers are formed over a gold electrode, the quantum rate principle, stated in Eq.~\ref{eq:nu}, can be directly compared to the traditional ET rate method, such as that proposed by Laviron~\citep{Laviron-1979}.

\section{Energy Degeneracy of Redox Reactions}\label{sec:degeneracy}

It has been demonstrated~\citep{Alarcon-2021} that if, in an ideal situation, the $\alpha$ symmetry factor of a diffusionless electrochemical reaction that follows the Butler-Volmer current-voltage equation~\citep{Bard-book}, is assumed to be 1/2, the ET constant, as proposed by Laviron~\citep{Laviron-1979} is simplified to

\begin{equation}
 \label{eq:k-Laviron}
	k_\alpha = \frac{e}{2k_BT}s,
\end{equation}

\noindent which, according to the presumption of $\alpha = 1/2$, implies that $k$ for anodic (oxidation) and cathodic (reduction) reactions are equivalent. Note that Eq.~\ref{eq:k-Laviron} was written considering the ET of an adiabatic single electron reaction, as for the situation depicted in Figure~\ref{fig:D-A}. To settle $k = G_0/C_q$ we assumed that the total number of states $N$ is unity in quantum rate theory equations.

In other words, $k = G_0/C_q$ corresponds to Eq.~\ref{eq:nu} with the consideration of $g_s$ degeneracy for an adiabatic ET of a single electron, leading to settle $G$ in Eq.~\ref{eq:k-zeroT} to be equivalent to $G_0$. This boundary condition implies simply to consider a single adiabatic ET ideal situation in which $\sum_{n=1}^{N}T_{n}\left( \mu \right) = N\exp \left(-\beta L \right) = N\kappa \sim 1$. Accordingly, for an adiabatic single ET, the resistance quantum is $R_q = 1/G_0 = h/g_se^2$, which is in a suitable agreement with the scheme proposed for a single electron transfer reaction depicted in Figure~\ref{fig:D-A}, as required. In other words, from the quantum thermal broadened ensemble, the individual states of the ensemble can be calculated provided that $N$ is experimentally accessible.

\begin{figure}[h]
\centering
\includegraphics[width=7cm]{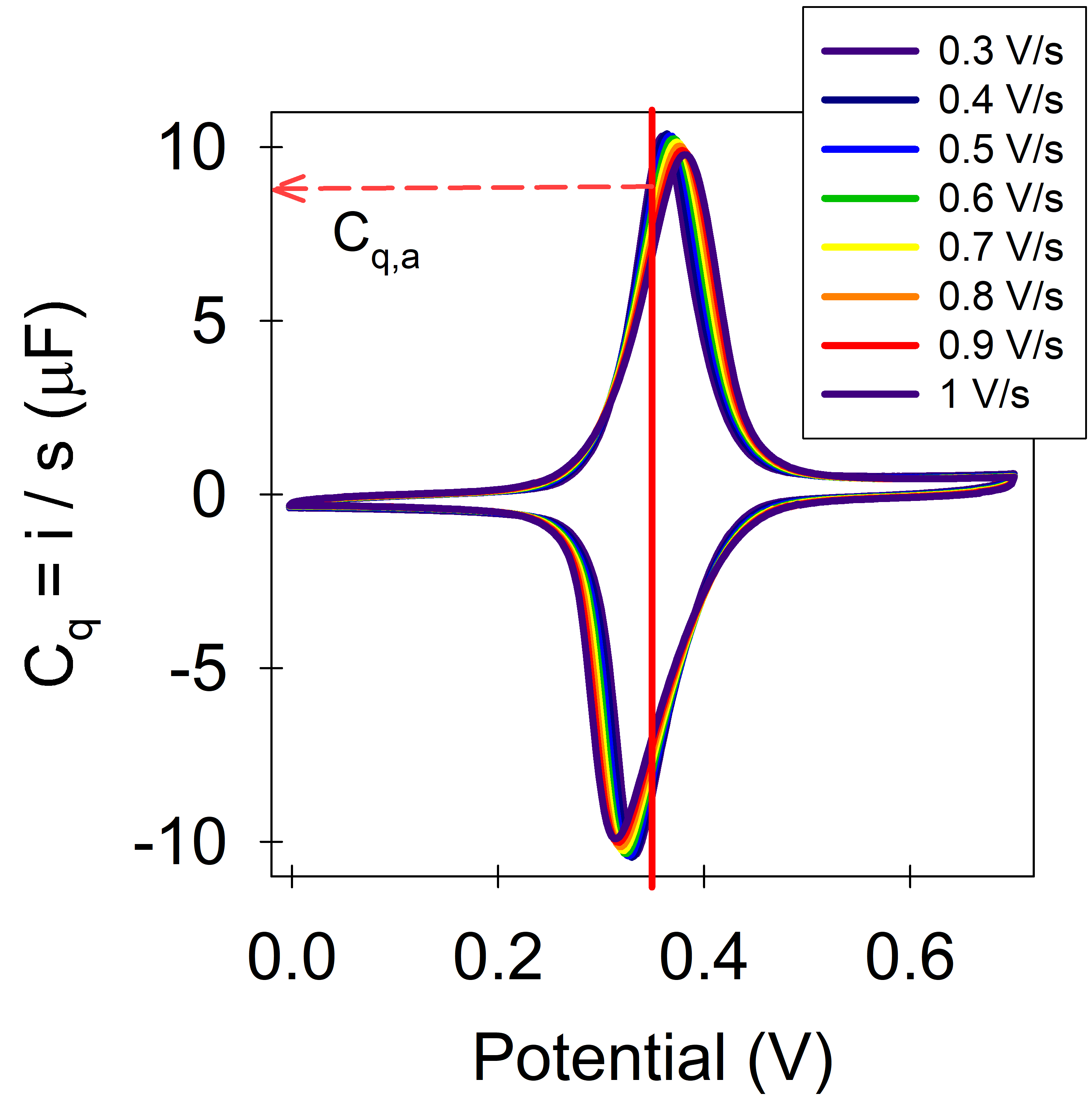}
\caption{Faradaic current associated with diffusionless ET reactions is defined in terms of $C_q$ as $i = C_q s$, where $s$ is the scan rate. For a dynamical electrochemical equilibrium, as is the case of well-behaved molecular redox films, $C_q$ is directly measured from a current-voltage curve as the ratio between $i$ and $s$, such that $C_q = i/s$. As indicated by a red arrow, the estimated value of $C_q$ at the formal potential is $\sim$ 8.5 $\mu$F. Hence, the rate $k$ of the ET reaction is $\sim$ 17 s$^{-1}$ and in agreement with Eq.~\ref{eq:k-gr}.}
\label{fig:k-QRate-Laviron}
\end{figure}

The proof that Eq.~\ref{eq:k-Laviron} is equivalent to $k = G_0/C_q = g_s e^2/hC_q$ is conducted similarly to the analysis performed in section~\ref{sec:Therm-QR}, that is, by considering the thermodynamics over the quantum rate dynamics. Hence, taking Eq.~\ref{eq:Cq-thermal} at the formal potential of EDOS junctions, where $f = 1/2$, the capacitance becomes $C_q = e^2/4 k_BT$ which results in an electric potential of $V = e/C_q = 4 k_BT/e$. Noting that the scan rate is $s = V/\tau = V k$, where $k = 1/\tau = 1/R_qC_q$ and substituting $V = 4 k_BT/e$ and rearranging, it leads to $k = \left( e/4 k_BT \right) s$, demonstrating that $k = k_\alpha/2$.

$k$ and $k_\alpha$ differ by an electrochemical degeneracy of $g_r = 2$. This degeneracy is incorporated in Eq.~\ref{eq:k-gr} and its origin will be discussed later in this section. Nonetheless, by taking $g_r = 2$ degeneracy into account, the quantum rate $k$ of Eq.~\ref{eq:k-gr} is equivalent to the rate $k_\alpha$ of Eq.\ref{eq:k-Laviron}. Hence, considering this $g_r = 2$ degeneracy, there is not only a theoretical agreement but experimental agreement between $k$ and $k_\alpha$~\citep{Alarcon-2021}. 

\begin{figure*}[h]
\centering
\includegraphics[width=14cm]{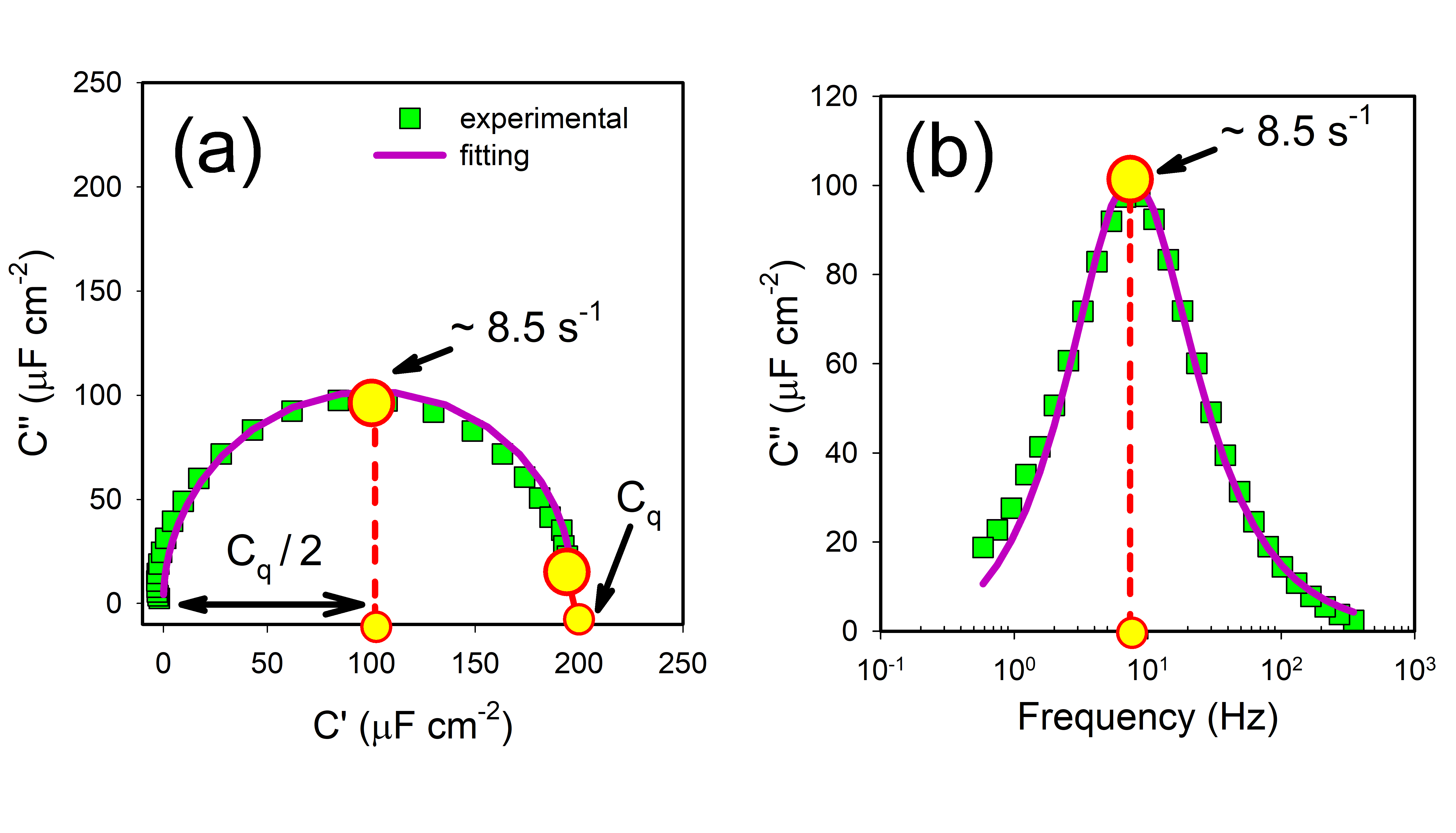}
\caption{(a) Nyquist capacitive diagram obtained from impedance spectra of an Fc-tagged monolayer. As the contribution of the non-faradaic response is minimal, as discussed in the text, the value of $C_q$ can be inferred directly from the diameter of the semi-circle, as indicated. Note that $C_q$ indicated in this Nyquist diagram corresponds to $C^0_q$ of Eq.~\ref{eq:Complex-Cq}. (b) Depicts a Bode capacitive diagram corresponding to the imaginary component of the complex capacitive response as a function of frequency. The frequency of the peak $\sim$ 8.5 s$^{-1}$ is a good estimation of $k$ if the degeneracy of $g_r = 2$ is taken into account, corresponding to an estimated $k$ of 17 s$^{-1}$, which is in excellent agreement with that obtained by the CV method. As the non-faradaic contribution is minimal, the total spectrum mostly reflects the contribution of the faradaic electrochemical response. The fitting to Eq.~\ref{eq:Complex-Cq} is excellent, as shown in this figure as a pink curve over the green squares.}
\label{fig:k-ECS-Laviron}
\end{figure*}

For instance, using Laviron's experimental approach, the anodic $k_\alpha$ was obtained as 19 $\pm$ 2 s$^{-1}$, whereas using $C_q$ value (obtained from Figure~\ref{fig:k-QRate-Laviron}) of $\sim$ 9 $\mu$F in Eq.~\ref{eq:k-gr}, the ratio $G_0/C_q$ is (155 $\mu$S)/(9 $\mu$F), providing a $k \sim$ 17 $\pm$ 0.9 s$^{-1}$, which is, within experimental error, in agreement with that obtained by Laviron's method. Note that $C_q$ is the only variable in Eq.~\ref{eq:k-gr}, since $g_r G_0$ is a constant with a value of $\sim$ 155 $\mu$S, which facilitates the estimation of $k$ in comparison with Laviron's approach. As shown in Figure~\ref{fig:k-ECS-Laviron}, $C_q$ value for the same Fc-tagged SAM can be also obtained as $\sim$ 8.6 $\mu$F, using impedance-derived capacitance measurements, resulting in a $k$ of 17 s$^{-1}$, which is in agreement with values obtained by cyclic voltammetry.

The consideration of a `redox' degeneracy state $g_r$ in the quantum rate, as noted in Eq.~\ref{eq:k-gr}, can be justified in two different ways. The first is by noting that redox reactions, at the formal potential of the electrode, are the result of two different and dynamical electric currents, owing to a dynamic equilibrium condition that leads to a resonant electric current. This type of resonant electrochemical current is a result of two types of charge carriers (electrons and holes), hence with two electrochemical currents, that is anodic (oxidation) and cathodic (reduction), respectively. The net resonant current is known as the exchange current $i_0$ (see complementary discussion in the next section). 

In terms of the `redox' degeneracy $g_r$, this is equivalent to considering an electron distribution probability at the resonance state of the electrode, meaning that initial and final fifty-to-fifty probabilities are established between reduced and oxidized states. In other words, this would correspond to multiplying the `existing probability' by a factor of two if two directions of ET are required to be taken quantitatively into account.

Alternatively, a second interpretation -- one that is complementary to the first way of interpreting $g_r$ -- is to consider the role played by the solvent/electrolyte, as indicated in Figure~\ref{fig:DA-electrode-electrolyte}, which implies that Ox + e $\rightharpoonup$ Red reaction occurring inside the monolayer is important. These internal redox reactions are discussed and interpreted in terms of 2D electronic characteristics resembling a 2DEG DOS. The existence of this type of redox dynamics was confirmed by quantum computational methods, as shown in the inset of Figure~\ref{fig:Fc-QMMM}. In this case, if the 2DEG electron-density is compensated by counter ions in the electrolyte, there will be an equivalent capacitance $1/C_\mu = 1/C_e + 1/C_q$ in the 2DEG structure as a consequence of the internal redox dynamics. This equivalent capacitance conducts to two superimposed `electrostatic' and quantum capacitive states in the redox sites, as depicted in Figure~\ref{fig:DA-electrode-electrolyte}. 

These two superimposed capacitive states of an individual redox site within the molecular layer in direct contact with the solvent/electrolyte environment shares the same elementary charge $e$ and possesses the same electric potential, i.e. $e/C_e \sim e/C_q$, conducting locally to $C_e \sim C_q$ in Eq.~\ref{eq:C-electroch}. Hence, the equivalent $C_\mu$ capacitance is $1/C_\mu = 2/C_q$, with a degeneracy of energy of $e^2/C_\mu = 2e^2/C_q$, where 2 is accounted as $g_r$ in the formulation of $k = g_r G_0/C_\mu$, as noted in Eq.~\ref{eq:k-gr}. This is equivalent to the previous electrical current degeneracy consideration of the origin of $g_r$ because $1/C_\mu = 2/C_q$ implies $C_q = 2C_{\mu}$. Taking that the redox capacitance of the interface as a result of the equivalent contribution of $C_e$ and $C_q$, within the redox centers of a molecular layer, there is an electrochemical current degeneracy for charging the capacitive states of the interface that is proportional to $C_\mu$ with an electric current $i = 2C_{\mu}s = (C_q + C_e)s = 2C_qs$ owing to the equivalent contribution of $C_e$ and $C_q$. Note that the equivalent $e^2/C_e$ and $e^2/C_q$ energy state degeneracy inherently conducts to two (anodic and cathodic) electrochemical currents contributing to the exchange net current $i_0$ of the interface.

Considering the origin of $g_r$ interpreted as a degeneracy of the electric potential in the redox sites (with the solvent contribution of $C_e$ to the electronic states $C_q$) is not only physically plausible, but it also provides a suitable understanding of the electric field screening of the solvent over the electron occupancy in molecules (occurring through $C_q$) during ET reactions, which is the essence of the effect of the solvent in ET reactions and of the origin of the super-capacitive phenomenon, which is intrinsically present in the field of electrochemistry and absent in the field of solid-state electronics. Note that this electric field screening phenomenon explains the role of solvent dynamics, including counter-ions besides the solvation. There is experimental support for this assumption, as stated in reference~\citep{Pinzon-2022}. The study of the appropriate electric field screening of the counter-ions over the electrodynamics of redox state sites that conduct to a $g_r$ energy degeneracy cannot only be applied to study ET reactions of monolayers, but it can be extended to study proton-coupled ET~\citep{Pinzon-2022} because the ion-electric field screening of the electron resembles the mechanism discussed in detail in reference~\citep{Pinzon-2022}.

Independently of the origin of the degeneracy, which still must be investigated in-depth, the value of $k$, for a well-behaved redox-tagged monolayer, can be obtained as~\citep{Alarcon-2021}

\begin{equation}
 \label{eq:k-gr}
	k = g_s g_r \left( \frac{e^2}{hC_q} \right) = g_r \left( \frac{G_0}{C_q} \right) = g_s g_r \left( \frac{E}{h} \right),
\end{equation}

\noindent corresponding to $k = 4 (e^2/hC_q) = 4 (E/h)$, in which the only variable to be experimentally determined is $C_{q}$. Therefore, once the degeneracy is taken into account, $C_q$ is the only variable required to determine $k$. The determination of $C_q$ as the only variable required to describe an electrochemical interface was theoretically predicted in reference~\citep{Bueno-2017-2}, using first-principle DFT to express the molecular origin of $C_q$. Using electrochemical methods, $C_q$ is measured from an ordinary electrochemical current-voltage scan or impedance-derived capacitive spectrum. In summary, because both $g_s$ and $g_r$ equals 2, $k$ is stated as $k = 4 (e^2/hC_q) = 2G_0/C_q = 4 (E/h)$, where $2G_0 = 4e^2/h$ is a quantum electrochemical conductance with a value of $\sim$ 155 $\mu$S. This constant value of 155 $\mu$S serves as a reference for calculating $k$ once $C_q$ is measured.

In the next section, it will be discussed the quantum mechanical nature of the charge transfer resistance.

\section{The Quantum Mechanical Meaning of the Charge Transfer Resistance}\label{sec:QRct}

Any discussion on the meaning of the resistance for the transport of electrons must start by formulating a expression for the carrier's velocity, as it is this term that defines the origin of the electric current $i$ that occurs under a potential difference of $V$. In the classical mode of electron transport in metals, the origin of $i$ charge variation in time $i = q/t$ is directly associated with the drift velocity $v_d$ of the carriers. Nonetheless, the origin of the charge transfer resistance $R_{ct}$, which is frequently measured in electrochemical experiments~\citep{Bard-book}, is not formulated or associated with the velocity of the carriers between the $D$ and $A$ species.

Nonetheless, the meaning of $R_{ct}$ in electrochemistry is taken from the Butler-Volmer equation, which defines the electric current in redox reactions as

\begin{equation}
 \label{eq:BV-equation}
	i = i_0 \left( \exp \left[ \alpha \frac{e}{k_B T}V' \right] - \exp \left[ \left( 1 - \alpha \right) \frac{e}{k_B T}V' \right] \right),
\end{equation}

\noindent in which $i_0$ is the exchange current, $\alpha$ the transfer coefficient and $V'$ the over potential (see below). In a diffusionless redox setting, as is the situation for heterogeneous electrochemical reactions wherein the redox entities are covalently coupled to the electrode (introduced in the preceding sections), the electric current associated with the meaning of $C_q$ (see Figure~\ref{fig:k-QRate-Laviron}) is obtained by perturbing the system around the formal potential $E_F/e$. Accordingly, the consideration of a very small over-potential\footnote{The over-potential is difference between the bias of potential $V$ and the formal (or Fermi level) $E_r/e$ potential of the electrochemical reaction.} $V' = V - E_F/e$ perturbation, close to the formal potential of the electrode reaction, permits to the expression of the charge-transfer conductance for a single electron transfer to be calculated as the derivative $di/dV$ of Eq.~\ref{eq:BV-equation}, which leads to~\citep{Sanchez-2022-1}

\begin{equation}
 \label{eq:Rct}
	G = \frac{1}{R_{ct}} = \frac{e}{k_B T}i_0,
\end{equation}

\noindent as a conductance that only depends on the electric current $i_0$ obtained at the Fermi level of the electrode, where the term $k_B T/e$ is a constant referred to as the thermal voltage.

The meaning of $R_{ct}$, as provided by Eq.~\ref{eq:Rct}, is directly correlated with Eq.~\ref{eq:k-gr} if the thermal voltage is defined as~\citep{Alarcon-2021, Bueno-2022}

\begin{equation}
 \label{eq:thermal-voltage}
		\frac{k_B T}{e} = \frac{s}{k},
\end{equation}

\noindent where $s = dV/dt$ is the scan rate or any type of time-dependent voltage perturbation.

According to Eq.~\ref{eq:Rct}, $R_{ct}$ can be rewritten as a function of $s$, such as $R_{ct} = s/ i_0 k$, in which $k$ is $G/C_q = 1/\tau$, where $\tau = R_q C_q$ and $R_q = 1/G$. Owing to $i_0 = C_q s$ (see Figure~\ref{fig:k-QRate-Laviron}), it conducts us theoretically to the conclusion that $R_{ct} = 1/G = R_q$, demonstrating that $R_{ct}$ is $R_q$. The above analysis not only leads to the conclusion that $R_{ct}$ is quantized, but also it provides meaning to $R_{ct}$ that is in agreement with the quantum rate theory.

The above theoretical examination was confirmed experimentally by analyzing the electrochemical interface using impedance spectroscopy. It is noteworthy that a classical analysis of the equivalent circuit of the interface, considers $R_{ct}$ to be separate from the solution and contact resistances, but this is not the case with quantum mechanical circuit analysis of the interface. A quantum mechanical interpretation of the resistance associated with electron transport in the redox reaction taking place at an interface (heterogeneous analysis) requires the resistance quantum to be the sum of all series resistances in the interface, which numerically complies with $1/G_0 \sim$ 12.9 k$\Omega$, as will be demonstrated. It is this limiting resistance that complies with the quantum mechanical interpretation of $k$ within the quantum rate theory, leading to the conclusion that the series resistance $R_s$ is part of a quantum interpretation of a charge transfer reaction that obeys a quantum limit of conductance.

For instance, a capacitance $C_q$ of $\sim$ 200 $\mu$F cm$^{-2}$ is estimated from a Nyquist capacitive plots as shown in Figure~\ref{fig:k-ECS-Laviron}\textit{a}, which is equivalent to $\sim$ 8.5 $\mu$F for an electroactive area of $\sim$ 0.044 cm$^2$. The calculated value of $\sim$ 8.6 $\mu$F is in agreement with values obtained in Figure~\ref{fig:k-QRate-Laviron} (a value of $\sim$ 8.5 $\mu$F). According to Eq.~\ref{eq:k-gr}, $k$ corresponds to $g_rG_0/C_q \sim$ 155 $\mu$S/8.6 $\mu$F $\sim$ 17 s$^{-1}$. Alternatively, $k$ can be obtained as twice (owing to the consideration of the $g_r$ degeneracy) the value of frequency $\sim$ 8.5 s$^{-1}$ obtained from Nyquist or Bode diagrams, as shown in Figure~\ref{fig:k-ECS-Laviron}, providing a $k$ of $\sim$ 17 s$^{-1}$.

Now, by fitting the spectra in Figure~\ref{fig:k-ECS-Laviron} to Eq.~\ref{eq:Complex-Cq} with the consideration of $R_q = R_s + R_{qt}$\footnote{The sum of $R_s$ and $R_{qt}$ correspond to the total series resistance of the interface and this is what has a quantum resistance limit, see more details in reference~\citep{Sanchez-2022-1}}, the value of $\sim$ 17 s$^{-1}$ is obtained as $k = g_r / [2\pi R_q C_q]\sim$ 17 s$^{-1}$, which provides the above-mentioned interpretation of the charge transfer resistance within a quantum limit of $\sim$ 12.9 k$\Omega$, confirming that an electrochemical reaction proceeds within a quantum efficiency of $G_0 = 2 e^2/h \sim$ 77.5 $\mu$S or $R_q = 1/G_0 \sim$ 12.9 k$\Omega$ per electron.

Alternatively, the theory can be verified by measuring the total conductance $G$, which is, according to the discussion conducted in section~\ref{sec:Therm-QR}, given by $G = \kappa G_0 N$, where $N$ is the total number of states. At the formal potential $N$ is calculated as $N = 4 k_BT C_q / e^2 \sim$ 6.3 $\times$ 10$^{12}$ and $\kappa = \exp \left(-\beta L \right)$. The average value $\beta$ can be considered to be $\sim$ 1.3 \AA$^{-1}$, and $L \sim$ 2 nm for the Fc-tagged monolayer with the impedance-derived capacitive spectrum measured in Figure~\ref{fig:k-ECS-Laviron}. For this experimental situation $G_0 = G/N\kappa \sim$ 77.5 $\mu$S, corresponding to $R_q = 1/G_0 \sim$ 12.9 k$\Omega$, which is in agreement with the theoretical value predicted by the quantum rate theory.

\begin{figure*}[h]
\centering
\includegraphics[width=14cm]{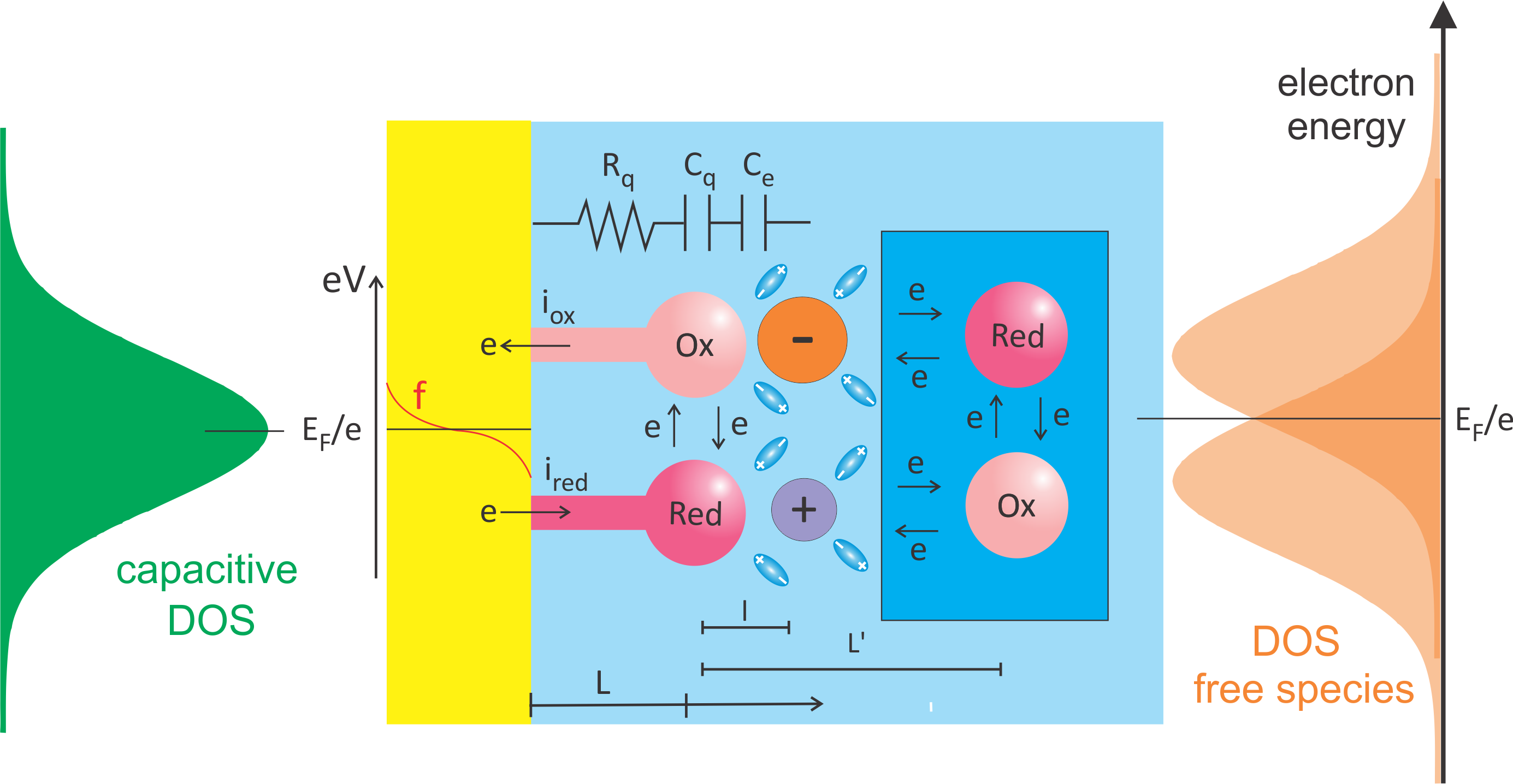}
\caption{Illustration of the quantum electrodynamics of redox reactions with an electric current flowing from the free $D$ species to and $A$ species (Ox/Red in the dark blue box) in the solution phase, intermediated by the presence of quantum capacitance states within the electrode. The intermediation of electric current flowing between the solution and the electrode by capacitive states does not imply the mass diffusion kinetic limit of the redox reactions as is always the case when metallic probe is in direct contact with free $D/A$ species in the solution environment. In the presence of capacitive states intermediating the redox reaction, a quantum resistance limit exists per electron transported in the interface, and this limit, as discussed in the text, is independent of $L$, $L'$, or $L + L'$ lengths.}
\label{fig:DOS-states-comparison}
\end{figure*}

Within the physical meaning of the quantum resistance (or conductance) that establishes that there exists a minimum limit of resistance (corresponding to a maximum of conductance) for the transfer of a single electron between $D$ and $A$ states in a redox reaction resides the quantum efficiency of biological processes wherein an electrolyte is intrinsically present to undertake the relativistic nature of the ET dynamics of redox reactions.

Undoubtedly, the quantum resistance value of $\sim$ 13 k$\Omega$ within an experimental error of 3.1\% matches with the theoretical value of $\sim$ 12.9 k$\Omega$. This not only has profound consequences for our understanding of electrochemical reactions within a relativistic quantum mechanics framework, but it also permits us to establish the quantum mechanics of biology~\citep{Bueno-2015}. Furthermore, as pointed out in reference~\citep{Alarcon-2021}, the quantum rate theory, when applied to study ET processes allows, not only studies unusual behavior of $\alpha$ symmetry factor but also permits separate kinetic regimes which are difficult to address in the framework of other quantum mechanical rate theories. For instance, in reference~\citep{Pinzon-2021} it has been used the quantum rate theory to study two different charge kinetic regimes operating in semiconductor thin films. In other words, as exemplified in this text, the quantum rate concept can be extended beyond redox monolayers to study different kinetic regimes and material compounds.

A suitable example will be provided in the next section, where it will be demonstrated that by appropriately modifying a metallic electrode with a monolayer containing redox moieties, a quantum transport can be achieved between the electrode and redox-free species in solution, providing a quantum channel path for electrochemical reactions that is independent of the distance between the redox moieties of the electrode and those redox species in solution/electrolyte bulk environment.

\section{Electron Transfer Mediated by Quantum Capacitive States}\label{sec:QR-mediation}

In this section, the application of the quantum rate theory to redox reactions in which electron transfer occurs in between the electrode and redox-free states in solution/electrolyte environment intermediated by quantum capacitive states within the interface of the electrode, as depicted in Figure~\ref{fig:DOS-states-comparison}, will be studied. 

It is well-known that the direct electronic communication of $D$ or $A$ states with metallic probes is kinetically limited by the transport of mass of $D$ and $A$ species from the solution phase to the interface of the electrode; that is, ET is limited by diffusion~\citep{Bard-book, Schmickler-book}. Nonetheless, whenever $D$ and $A$ species are covalently attached to the electrode, the electronic communication of redox states with the electrode is no longer limited by diffusion and the faradaic ET electric current $i$ is directly proportional to $C_q$ by a time-dependent potential perturbation $s = dV/dt$, such that $i = C_q s$, as it was discussed in the previous sections (see Figure~\ref{fig:k-QRate-Laviron}).

Accordingly, it is important to scrutinize the mechanisms of electron transport and electric communication between free redox states in the solution/electrolyte bulk phase environment with that of the electrode in the presence of capacitive states in the electrode that are capable of mediating the electron transport from free $D$ and $A$ species (in solution) to the electrode using $C_q$ states, as indicated in Figure~\ref{fig:DOS-states-comparison}. The equivalent circuit analysis can be conducted in the same manner as that discussed in section~\ref{sec:QRct}, from which a limiting resistance was evaluated for the faradaic quantum ET branch of the circuit indicated in Figure~\ref{fig:circuit-azurin}. More details can be found in work of Sanchez and colleges~\citep{Sanchez-2022-2}.

\begin{figure}[h]
\centering
\includegraphics[width=7cm]{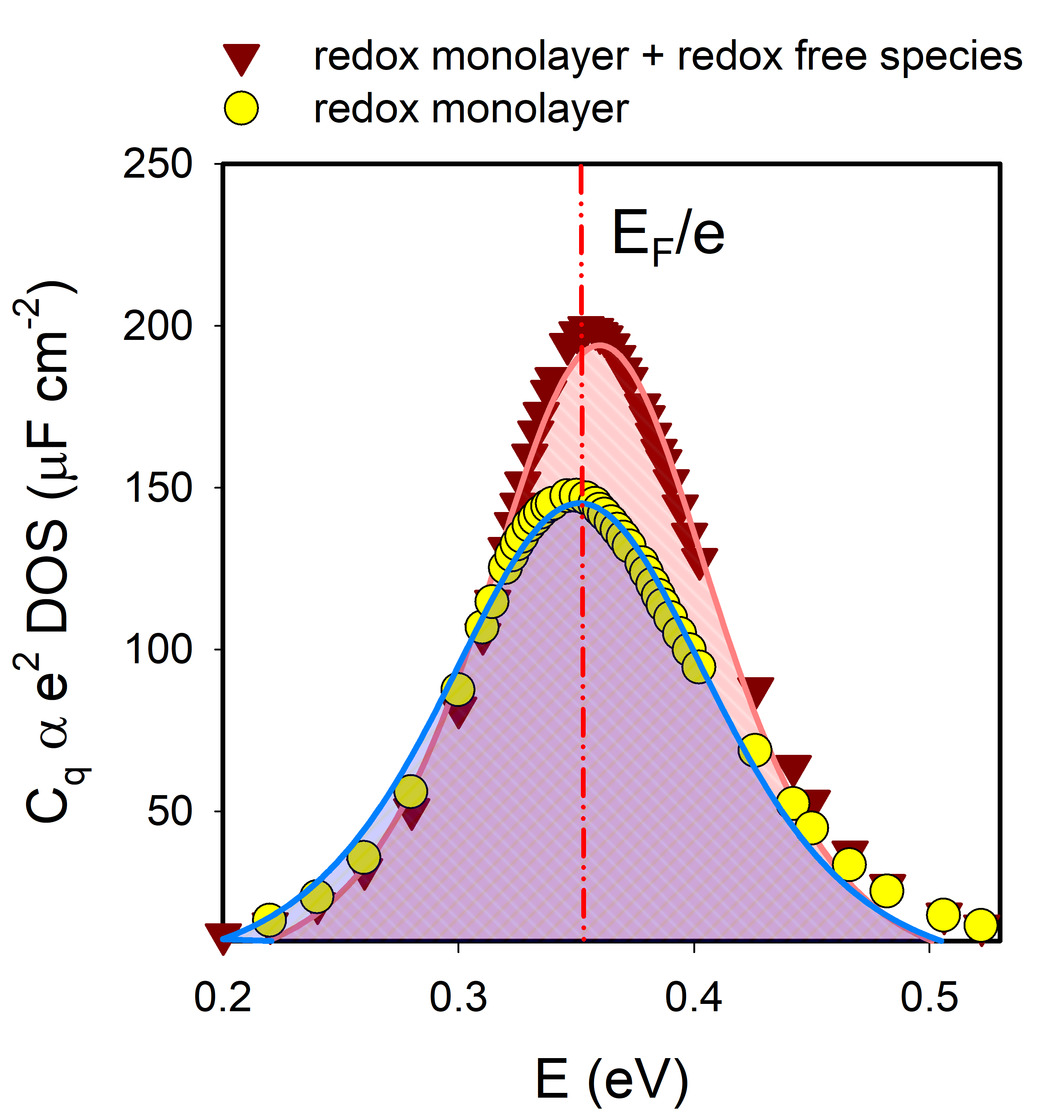}
\caption{EDOS of a redox molecular film in the absence (yellow dots) and presence (pink dark triangles) of $D/A$ free species in the solution phase. As the formal potential $E_F/e$ of $D/A$ free species is in alignment with that of capacitive states of the monolayer, the EDOS keeps its shape and the only role of $D/A$ free species is to increase the quantum modes of electron transport or the quantum-capacitive energy states of the interface. The fitting of both EDOS is in good agreement with the theory, i.e., with Eq.~\ref{eq:Cq-thermal}.}
\label{fig:DOS-superimposed}
\end{figure}

As the shape of a redox DOS chemically coupled to an electrode can be measured, as reported in section~\ref{sec:ChargingElecInterfaces}, possessing a Gaussian-like shape that is essentially governed by the $f(1-f)$ term of a thermally broadened capacitive function $C_{q} = \left( e^2N/{k_{B}T} \right) f(1-f)$, it is possible to compare the DOS of the redox film in the absence and the presence of free Ox/Red states in the solution/electrolyte bulk phase. 

A similar profile exists between the DOS measured in the presence and absence of free Ox/Red states in solution/electrolyte, as shown in Figure~\ref{fig:DOS-superimposed}. Both DOS shapes have a maximum at $E_F/e$, corresponding to the Fermi level of the electrochemical junction, as shown in Figure~\ref{fig:DOS-states-comparison}. The only difference observed between the DOS in the absence or presence of free Ox/Red states in the solution/electrolyte phase is the peak height at the maximum. The higher peak height in the presence of free Ox/Red states in solution indicates an increment in the population of the DOS and consequently of $N$ quantum channel modes~\citep{Sanchez-2022-2}. 

The conclusion is that the presence of free Ox/Red states in the solution/electrolyte does not change the shape of the capacitive DOS function. Hence, the presence of free Ox/Red states in solution/electrolyte does not influence the energy distribution of the states, but it increases the population of states, with differences in the DOS interpreted as an increase in the $N$ number of quantum channels. This interpretation of a higher $N$ is confirmed by studying the Gaussian-like DOS shapes measured in the presence and absence of free Ox/Red states in the electrolyte, as shown in Figure~\ref{fig:DOS-superimposed}. It can be observed that these shapes have an equivalent standard deviation of $\sim$ 0.4 mV and a similar position of the maximum~\citep{Sanchez-2022-2}, confirming that these shapes are equivalent with differences owing to an increase in the population $N$ of the states.

Additionally, the study of the quantum rate dynamics conducted in the presence of free Ox/Red states in solution/electrolyte lead to the conclusion that there is a quantum resistance limit of $R_q$ $\sim$ 12.9 k$\Omega$. This resistance quantum limit for the diffusionless ET dynamics established in the interface was observed to be independent of the backbone structure of the redox monolayer and of the presence of free redox states in the electrolyte environment. In other words, the rate efficiency with which electrons of free Ox/Red couple moieties in the electrolyte communicates with the electrode (at the formal potential level), intermediated by quantum Ox/Red capacitive states of the monolayer, follows particular quantum mechanical rules. These rules comply with the quantum rate theory, which predicts a resistance quantum of $R_q$ $\sim$ 12.9 k$\Omega$ for electrons; this limit is established in the absence or presence of a redox probe in the solution~\citep{Sanchez-2022-2}. 

The limiting value of $R_q$ is adjusted by the rate $k$ in the interface and is measured independently of variations in the value of $C_q$ for different monolayers. This suggests that the number of steps in which ET is conducted does not influence the efficiency of charge transport. Therefore, the value of $k$ varies (or adjusts) according to different values of $C_q$ obtained in each interface, conducting to a $R_q$ limit per each individual electron transfer, which allows a maximum $G = 1/R_q$ per electron~\citep{Sanchez-2022-2}, demonstrating the maximum efficiency for the transport of the electron, which complies with the relativistic nature of the ET reaction. In summary, the analysis suggests that electrochemical reactions apparently use the environment to conform towards a maximum quantum efficiency for the ET reaction to proceed.

The quantum rate theory is significantly important for understanding the efficiency of ET between electrochemical species in solution and electrodes even when the reaction is intermediated by molecular states. Following the quantum rate model, it is possible to state that the ensemble of molecules is behaving as a quantum-capacitive energy state level at the interface, which can be aligned with the formal energy of the redox species in the solution phase, allowing us to control and align the energy levels of ET reactions~\citep{Sanchez-2022-2}. The alignment of the energy level of a redox-modified electrode with redox-free species in solution permit a higher electrochemical current to be established as a consequence of a higher $N$.

The most important conclusion achieved so far from a theoretical and experimental analysis is that a maximum efficiency of electron transport (within a resistive quantum limit of $\sim$ 12.9 k$\Omega$) is observed if there is alignment between the formal potential states. This resistive limit for ET reactions is achieved independently of the length $L$, $L'$ or $L + L'$, as particularly investigated (see Figure~\ref{fig:DOS-states-comparison}) during the interfacial ET processes mediated by capacitive redox states ~\citep{Sanchez-2022-2}. 

This limiting value is obtained from the analysis of $G_0 = G / N \kappa$, where $R_q = 1/G$ is the total series resistance of the interface (comprising solution and contact resistances). The conclusion is that the electron conductance associated with ET reactions is quantized by amounts of $G_0$, as confirmed by a series of experiments in which $G$ is measured and correlated with $C_q$ by calculating $N$. Note that the normalization of $G$ by $N$ and presuming that $\kappa \sim \exp(-\beta L)$ is available or can be estimated, it allows us to calculate the total conductance $G$ as a sum of quantum modes of conductance in the interface. This analysis permits us to validate the use of quantum rate theory to model ET transfer reactions.

The quantum rate theory permits a quantum mechanical analysis of redox reactions using a simple experimental set-ups. The analysis allows us to conclude, as was the initial hypothesis, that electron transport follows a relativistic $E = h\nu$ wave dynamics with a rate of $\nu = e^2/hC_q$ intrinsically related to the ET dynamics. The ET rate that governs the kinetics of the electrochemical reaction is adjustable to contact and the electrolyte environment such that it conforms to a maximum efficiency for the transport of electrons, in agreement with the hypothesis stated in section~\ref{sec:introduction}, where the quantum mechanics of ET reactions was predicted and was further demonstrated to follow relativistic wave dynamics. This conducts to a minimum loss of energy, for the transport of the electrons, independently of the length of the quantum channel because the environmental conditions are adjustable.

It is noteworthy that the relativistic dynamics of ET reactions resembles the transmittance of electromagnetic waves through space. This can be comprehended (or interpreted) through a time-induced charge-variation analysis based on an individual $n$ quantum channel of length $L$ with a given density of $(dn/dE)$. If it is considered a Fermi velocity for electrons to travel in the channels of $c_*$, it permits modelling of the changes in the electrochemical faradaic current as

\begin{equation}
 \label{eq:electric-current-Qrate}
	\delta i = -e \left( \frac{c_*}{L} \right) \delta \mu \left( \frac{dn}{dE} \right),
\end{equation}

\noindent from which $G$ is obtained by substituting $\delta \mu = -e \delta V$ in Eq.~\ref{eq:electric-current-Qrate}. Further rearranging leads to

\begin{equation}
 \label{eq:G-v_F-DOS}
	G = \frac{\delta i}{\delta V}  = e^2 \left( \frac{c_*}{L} \right) \left( \frac{dn}{dE} \right).
\end{equation}

\noindent from which, by noting the definition of the DOS for a perfect quantum channel as $(dn/dE) = g_s L/c_*h$ and substituting it in Eq.~\ref{eq:G-v_F-DOS}, it is obtained that $G = g_s e^2/h = G_0$, which can be simplified as

\begin{equation}
 \label{eq:G-v_F}
	\frac{G}{C_q} = \left( \frac{c_*}{L} \right),
\end{equation}

\noindent which is equivalent to Eq.~\ref{eq:nu} whenever $g_s$ is considered in the definition of $\nu = e^2/hC_q$ rate. This is an expected result based on the relativistic quantum rate description of a single and ideal ET step with a spin degeneracy of $g_s$, as was the initial presumption of section~\ref{sec:QrateT+EChem}.

For the sake of simplicity in the physical analysis of the problem, let us consider adiabatic ET where $\kappa$ is unity. In this case, the meaning of $c_*/L$ is settled by noting that $L = n\lambda$, where $n$ is the number of quantum state modes within $L$ for the quantum transmittance (see Figure~\ref{fig:Qchannel}), i.e. $nc_*/\lambda$, from which it can be demonstrated that each quantum channel $n$ operates independently. 

Implicit in the analysis of $G/C_q = c_*/L$ is that if $G$ is normalized by $N$ (the total number of channels $n$), which is obtained in the Fermi level of the ET reactions, the corresponding situation is $G_0/C_q = c_*/\lambda$. The latter is owing to $G_0 = G/N$ for the adiabatic ET situation. This particular setting is provided by Eq.~\ref{eq:Cq} which, as theoretically and experimentally discussed in this text, corresponds to a $C_q$ that is proportional to the DOS of the redox film assembled over the electrode, from which $N$ can be easily obtained. Note that $C_q$ contains all the required information for the analysis of ET reactions, as predicted by~\citep{Bueno-2017-2}.

Within Eq.~\ref{eq:G-v_F} is also the meaning of $i_0$ as a function of $c_*$. Section~\ref{sec:QRct} started by noting that any definition of resistance must be correlated with the velocity of the carriers. Hence, noting from Eq.~\ref{eq:Rct} that the thermal voltage for a single adiabatic electron transfer (where $N$ and $\kappa$ equates to unity) is $k_BT/e = i_0/G$ at the Fermi level (where $f = 1/2$) and from Eq.~\ref{eq:Cq-thermal} $k_BT/e$ is equivalent to $e/4C_q$, it is possible to demonstrate, taking into account the $g_r$ energy state degeneracy, that $i_0$ is

\begin{equation}
 \label{eq:i0-general}
	i_0 = \left( \frac{g_r e}{4} \right) \left( \frac{G_0}{C_q} \right) = e \left( \frac{e^2}{hC_q} \right),
\end{equation}

\noindent where the value of 4 will be compensated by the product of $g_s = 2$ and $g_r = 2$. The case of a single adiabatic ET is implying that $L = \lambda$. Noting Eq.~\ref{eq:G-v_F} for this situation, Eq.~\ref{eq:i0-general} can be rearranged as

\begin{equation}
 \label{eq:i0}
	 \nu = \left( \frac{e^2}{hC_q} \right) = \left( \frac{c_*}{\lambda} \right) = \frac{i_0}{e},
\end{equation}

\noindent which not only establish a direct correlation between $i_0$ and $c_*$, as is required for an appropriate physical description of the meaning of charge transfer resistance $R_{ct} = 1/G_0$ of a single electron in the field of electrochemistry, but Eq.~\ref{eq:i0} also establish the meaning of $i_0$ directly as a function of $\nu$ through the elementary charge of the electron $e$. Noteworthy is that $i_0 = e\nu$, where $i_0 \propto c_*$ is established as $e/\lambda$, which is the electron density in the quantum channel. 

The latter proportionality is intrinsic to the relativistic nature of redox reactions as depicted in Figure~\ref{fig:D-A}. This relativistic nature implies that $i_0$ of a single electron is ambipolar, meaning that an electric current is formed for one electron and one hole as a charge carrier in opposite direction. The physical meaning of the latter statement is that there is the transport of one single electron and one single hole carrier in opposite directions whereas the chemical meaning is defined in kinetics terms, that is, the oxidation and reduction electric currents are of the same magnitude but in the opposite direction and the electrochemical reaction is equilibrium.   

It is particularly interesting that redox reactions occur at frequencies between 100 Hz down to 0.1 Hz, which corresponds to a frequency region of the electromagnetic spectrum classified as extremely low frequencies, that is, a region in which the wavelength is extremely high. Frequencies in this region permit a multitude of possibility for $\nu \sim c_*/\lambda = i_0/e$ in agreement with Eq.~\ref{eq:G-v_F} and~\ref{eq:i0}. These possibilities are due to the high degree of freedom that can be achieved for the ET dynamics thanks to the electrolyte environmental. 

In other words, the electrolyte conformation allows ET to occurs at frequencies that guarantee the maximum efficiency of electron transmittance (corresponding to the quantum limit of electron transport) in different electrolyte settings (ion and counter ions distances) and solvent environments (solvating conditions). The adjustment of the ET frequency for the maximum efficiency of electron transport is intrinsic to the high efficiency with which electricity is conducted in biological systems, a condition that cannot be easily attained using solid-state electronics. 

Note that this phenomenon also corresponds to frequencies close to that in which electricity is transmitted by transmission lines power stations using the AC mode of transport, a frequency and wave mode of transmittance that is required to decrease the loss of energy during the transmission of the electrical energy at long distances. However, the equivalence is complete here because ET processes are governed by relativistic quantum mechanics whereas the transmission of electric energy (using metallic contacts) is classical. 

Therefore, the velocity $c_*$ of the electron transmittance in Eq.~\ref{eq:G-v_F}, according to relativistic quantum mechanics, cannot be estimated from a fixed $L$ owing to $L \sim \lambda$ (for a single adiabatic state mode of electron transmittance), but it can theoretically be varied from lower values, such as 10$^{-3}$ m s$^{-1}$ (the drift velocity in metals) up to the Fermi velocity $c_*$ which is $\sim$ 10$^6$ m s$^{-1}$. This permits electrons to `hop' from one $D$ to another $A$ state with a maximum relativistic quantum mechanical efficiency that depends on the $\nu = e^2/hC_q$ rate that permits electrons to be transported in different chemical environments comprising long distances if required, as is the case of the respiration chain in biological systems~\citep{Bueno-2015}, for instance. 

The quantum capacitive mediation of the electron transfer reaction demonstrates an improvement in electron transport between the electrode and redox free species in solution, which cannot be achieved by the direct contact of a metal with redox-free states in the solution/electrolyte bulk phase. The mediation of ET reactions using quantum capacitive states can effectively improve the electron transfer to a quantum efficiency whether $C_q$ states are intermediating the transport. This quantum efficiency is obtained independently of a non-adiabatic bridge within an electron coupling of $\kappa \sim \exp \left( -\beta L \right)$\footnote{Note that for the transfer of a single electron through a barrier of length $L$ within a single quantum channel mode this becomes $\exp \left( -\beta \lambda \right)$ as $L = \lambda$, in agreement to the scheme of Figure~\ref{fig:Qchannel}.} or contact and solution resistances. The resistance that obeys quantum rules in the interface is the series resistance, where it includes contact- and solution-phase resistance. It is a resistance phenomenon that has been referred to and interpreted in the past as the uncompensated resistance~\citep{Bueno-2013} of the interface. 

In summary, it is possible to achieve quantum mechanical efficiency for the transport of electrons from the electrolyte to the electrode if the interface of the electrode is chemically modified with redox capacitive states that serve as intermediate energy levels. The use of classical mechanics reasoning for the latter assumption is counterintuitive, and ET reactions in the latter context are performed under a diffusionless condition. Accordingly, molecularly-modified metallic electrodes have proven to be better for probing redox reactions, permitting a better quantum mechanical dynamics for the flowing of electrons from a solution phase redox states towards the electrode and \textit{vice-versa}. Direct metallic contact between the electrode and redox-free species in solution is not appropriate for probing the quantum electrodynamics of redox reactions. Consistently, the presence of $C_q$ states intermediating the reaction permits electrons to be transmitted with a charge transfer resistance that complies with $\sim$ 12.9 k$\Omega$, leading to a maximum electrode-mediated quantum rate efficiency. This phenomenon can only be understood from a relativistic quantum mechanical perspective of electron transport.

\section{Final Remarks and Conclusions}

It can be concluded that the electrodynamics of the redox reactions follows a fermionic dynamics that such as it is observed for graphene~\citep{Bueno-2022} and Weyl semi-metals can only be appropriately described using relativistic quantum mechanics.

Here, a first-principle fundamental quantum rate $\nu = e^2/hC_q$ concept was used to elucidate the electron transfer rate constant $k$ of electrochemical reactions as $k = G/C_\mu$~\citep{Bueno-book-2018}, in Eq.~\ref{eq:k-Celect-zeroT}, where $G$ and $C_\mu$ have general definitions. $G$ was settled according to the Landauer definition of Eq.~\ref{eq:Landauer}, whereas the $C_\mu$ of Eq.~\ref{eq:C-electroch} was stated as a series combination of $C_e$ and $C_q$ components, i.e. $1/C_\mu = 1/C_e + 1/C_q$. Hence, Eq.~\ref{eq:k-finiteT} permitted the electron rate constant of faradaic reactions to be established as $k = G/C_q$, which is a particular setting of $k = G/C_\mu$ (Eq.~\ref{eq:k-Celect-zeroT}) in which only faradaic charging dynamics is considered. This faradaic dynamics follows relativistic quantum mechanical rules that conforms with the definition of $\nu = e^2/hC_q$.

Therefore, although a particular $k_{nf} = G_{nf}/C_e$ concept is also able to be employed to access the dielectric-polarization dynamics within an electrochemical interface to study particular `double-layer' type of electrochemical dynamics that does not involve the transfer of electrons, the faradaic $k \propto e^2/hC_q$ component is of particular importance to model redox reactions. The faradaic dynamics in redox-active monolayers can be entirely separated from the dipole-dielectric dynamics using impedance-derived capacitance spectroscopic methods. This method permits the investigation of the faradaic separately from the dielectric polarization dynamics in great resolution and detail.

The theoretical analysis focused on the meaning of $C_q$ and $\nu \propto e^2/hC_q$ demonstrates that homogeneous Marcus-Arrhenius semi-classical ET theory is solely a particular thermodynamics setting of the above definition of $k = G/C_q$, where $G \propto \kappa G_0 N$ and $\kappa N = \sum_{n=1}^{N}T_{n}\left( \mu \right)$ is the transmission probability of all $N$ quantum modes of transmittance. 

As part of Marcus semi-classical ET theory, heterogeneous diffusionless ET dynamics were studied as a particular case of $k = G/C_q$, where particular boundary conditions were determined for resolving $G$ and $C_q$. For instance, for molecular films, where redox centers can be well arranged with a well-defined and controlled $L$ distance from an electrode, it is possible to model the quantum conductance $G$ of these molecular films as $G = \kappa G_0 N$, where $N = L/\lambda$ is interpreted as the total number of quantum states (or quantum channels) for the transport of the electrons between the redox moieties and the electrode. 

As the redox moieties are estimated to be within a fixed length $L$ from the electrode, redox films containing channels within an average length $L$, permit ET dynamics to be represented solely in terms of $G_0$ (the conductance quantum constant $\sim$ 77.5 $\mu$S) and $\kappa \sim \exp \left( \beta L \right)$ (the electronic coupling). The electronic coupling is a constant value in this setting and permits the evaluation of a qualitative analysis of the ET if required. If $\kappa$ is available the ET reaction can be evaluated quantitatively.

The electrochemical capacitance permits to access the energy of the ET dynamics and is interpreted to be degenerated, i.e., it is related to $g_s$ and $g_r$ degeneracies such as $E = (g_s g_r)(e^2/C_q)$. This energy, for ET reactions, is established at room temperature and hence requires the statistical mechanics consideration that, for heterogeneous reactions, follows Eq.~\ref{eq:Cq-thermal}. Accordingly, at the Fermi level of the electrode, $f$, in Eq.~\ref{eq:Cq-thermal}, is 1/2, permitting the number of quantum states $N$ to be obtained simply as $N = 4 k_BT C_q/ e^2$. 

Since $G$ and $C_q$ are experimentally accessible, it is quite straightforward to obtain $N$ and all required parameters to investigate the relativistic electrodynamics of ET reactions in different nanoscale settings using an electrode as a probe. Additionally, within estimations of $L$ and $\beta$, which are required for knowing the electronic coupling constant of the potential barrier established between $D$ and $A$ states, all the parameters of the ET reactions are obtainable to quantitatively study the relativistic electrodynamics of ET reactions using quantum rate theory. This experimental and theoretical approach permits the investigation (qualitatively or quantitatively) of the relativistic quantum electrochemistry of ET reactions in great detail, as demonstrated in this study.

The advantage of the quantum rate to describe ET constant compared to previous ET approaches is that it not only provides new physical insights correlating nanoscale and molecular electronic concepts such the conductance quantum $G_0$ (developed within a physical science perspective) with the ET rate constant (developed within a chemical science perspective) through an experimentally measurable quantum capacitive $C_q$ concept, but it also permits to investigate the quantum electrodynamics involving these concepts with great accuracy. For instance, previous theoretical ET approaches were developed exclusively focused on describing homogeneous ET reactions, a setting that restricts the study of ET processes that requires an electrode to probe the ET reaction.

In summary, the adaptation of the homogeneous ET setting to describe the heterogeneous ET situation has great limitations, not only because it is not constructed using the first principles of quantum mechanics but because the parameters of the model are inferred and cannot be directly compared with experiments, and hence the comparison of experiments is hard and limited. The present quantum rate theoretical approach not only comprises previous theoretical models in their specific settings (e.g., homogeneous ET reaction) but also permits the investigation of heterogeneous ET reaction conditions, in which all the parameters of the model can be experimentally measured using time-dependent methods such as impedance-derived capacitance spectroscopy. The ability to test the theory through experiments constitutes the major advantage of the present introduced theoretical quantum rate approach over the previous one and can be further explored to study different kinetic regimes, which are difficult to address in the framework of previous quantum mechanical rate theories.

\section{Credit author statement}

Paulo R. Bueno: Conceptualization, data curation, writing the original and final drafts, picture-writing and editing, funding acquisition.

\section{Acknowledgments}

The author is grateful to the Sao Paulo State Research Foundation (FAPESP) for grants 2017/24839-0 and to the National Council for Scientific and Technological (CNPq).







\printcredits

\bibliographystyle{cas-model2-names}

\bibliography{biblio}

\end{document}